# Kinetics of the surface-nucleated transformation of spherical particles and new model for grain-boundary nucleated transformations

Nikolay V. Alekseechkin

Akhiezer Institute for Theoretical Physics, National Science Centre "Kharkov Institute of Physics and Technology", Akademicheskaya Street 1, Kharkov 61108, Ukraine
Email: n.alex@kipt.kharkov.ua

KJMA theory; Avrami exponent; grain boundaries; size distribution.

Equations for the transformed volume fraction of a spherical particle with nucleation on its surface, both nonisothermal and isothermal, are derived in the framework of Kolmogorov method adapted for this problem. Characteristic parameters governing the transformation kinetics are determined; the latter is studied with particular emphasis on the Avrami exponent temporal behavior. It is shown that the surface-nucleated transformation qualitatively differs from the bulk-nucleated one at large values of the characteristic parameters due to the one-dimensional radial growth of the new phase occurring after the complete transformation of the surface itself at the early stage of the process. This effect also manifests itself in the considered ensemble of size-distributed particles and in the grain-boundary nucleated transformations. The logarithmic normal distribution inherent for the particles obtained by grinding is employed for numerical calculations and shown to stretch temporally the volume-fraction and Avrami-exponent dependences for the ensemble of identical particles. A new model for grain-boundary nucleated transformations alternative to the Cahn model of random planes is offered; it is based on the ensemble of size-distributed spherical particles with the possibility for a growing nucleus to cross grain boundaries. The kinetics of this process is shown to be governed by the same characteristic parameter, as for a single particle, and qualitatively differs from the Cahn-model one. In particular, the logarithmic volume-fraction plot at large values of the governing parameter ends by a characteristic bend observed on experimental curves for the crystallization of bulk metallic glasses. This peculiarity together with the form of the plot as a whole directly indicates to the grain (polycluster) structure of metallic glasses and nucleation at intercluster boundaries.

## 1. Introduction

The kinetics of a phase transformation process in infinite space under nonisothermal conditions is described by the following equation for the volume fraction (VF) $X_K(t)$ of the material transformed:

$$X_K(t) = 1 - \exp\left[ -\int_0^t I_b(t') V(t',t) dt' \right], \quad V(t',t) = gR^D(t',t), \quad R(t',t) = \int_{t'}^t u(\tau) d\tau \quad (1a)$$

where $I_b(T(t))$ and $u(T(t))$ are the bulk nucleation rate and the velocity of nucleus linear growth, respectively, which are functions of temperature $T$; $g$ is the geometrical factor and $D$ is the space dimensionality. This equation was derived by Kolmogorov in the framework of rigorous probabilistic method under certain model assumptions; an English translation of this remarkable work is available [1].

Under isothermal conditions ($I$ and $u$ are constant), the integral is calculated giving

$$X(t) = 1 - \exp(-k(T)t^n), \quad k(T) \equiv (g/n)I(T)u(T)^D, \quad n = D + 1 \quad (1b)$$

This equation was derived by Kolmogorov [1] from general Eq. (1a) in two variants: for continuous ($n = 4$) and instantaneous ($n = 3$, $k = (4\pi/3)N_v u^3$, $N_v$ is the volume density of centers) nucleation. Despite the fact that the case of diffusion-limited growth is beyond the Kolmogorov model, the high accuracy of Eq. (1b) was shown for this case also [2]; $V(t',t) \sim (t - t')^{D/2}$ results in $n = D/2 + 1$.

Johnson and Mehl (JM) [3] have solved the difficult problem of finding $X(t)$ quite differently, offering an inventive method based on the fictitious mechanism of nucleation and growth; this approach includes the so-called "phantom nuclei" into consideration. Although JM's approach is valid in a nonisothermal case also [2] (under the assumptions of Kolmogorov's model) and Eq. (1a) can be obtained within this approach (cf. Appendix), the authors considered the isothermal case and arrived at Eq. (1b) with $n = 4$; at the same time, the particular nonisothermal case with the linear time dependence of $I(t)$ and $u(t)$ was considered as well. The full text of this classical work including important Appendices (unpublished earlier) is kindly presented in ref. [4]. Being an inherent part of the ensemble of nuclei [2, 5, 6], the phantom ones create serious difficulties in calculating VFs in the cases beyond the Kolmogorov model, e.g., when several phases simultaneously nucleate and grow with different velocities or when a one phase grows by a diffusion-type law; a new method for solving this issue was offered and the effect of phantom nuclei in both these problems was estimated in refs. [2, 7].

Avrami assumed quite different nucleation mechanism – nucleation by "germ nuclei" which already exist in the old phase and whose density diminishes during the transformation [8]; in modern terminology, this is continuous nucleation at pre-existing centers. Eq. (1b) with $n = 4$ and $n = 3$ gives two limits for this type of nucleation corresponding to small and large values of the quantity $I_c t$, where $I_c$ is the specific nucleation rate at a centre [9-11]. In this way, Avrami considered Eq. (1b) as a general one for phase transformations, where $n$ can vary between the mentioned limits as well as for other geometries [9]; since, $n$ is commonly called the Avrami exponent (AE).

In view of the great importance of the KJMA theory for materials science and other applications, its development and extensions continue the last decades [2, 5-7, 11-21]; in particular, this concerns the studies of the issue of grains size distribution [22-26]. Eqs. (1a, b) are suitable for transformations proceeding in sufficiently large samples and lose their validity for ones occurring in small objects such as a powder particle or a droplet. Equations replacing Eqs. (1a, b) in the case of a finite spherical domain were derived in ref. [11] for the case of bulk nucleation. However, experimental observations show that nucleation can occur on the surface of a particle in the processes of crystallization of both amorphous powders [27] and liquid droplets [28]. Therefore, the study of kinetics of the surface-nucleated crystallization of a spherical particle and an ensemble of size-distributed particles is topical and the present report is dedicated to solving this problem. The original Kolmogorov method is adapted for this purpose and the procedure of solving is quite similar to that of ref. [11]. The AE behavior during the transformation process is of particular interest here. The results can be useful, in particular, for studying the processes of crystallization of amorphous powders [29, 30].

Another important application of VF equations for an ensemble of size-distributed particles is modeling the grain-boundary nucleated transformations by them. Phase transformations originating from grain boundaries in polycrystals [31] or polycluster amorphous solids [32] often occur. The classical Cahn model [33] represents the network of grain boundaries by the set of random planes. The advantage of this model is in that it gives an exact and comparatively simple solution for the VF, however, there is the issue of its adequacy to the real network and the corresponding transformation kinetics: (i) faces of numerous adjacent grains are flat and lie in a one plane; (ii) small angles between faces occur; (iii) after the "saturation" of boundaries [33], when they are covered entirely by the new phase, the space filling proceeds by the 1D growth of infinite flat plates and is described by the corresponding long-time asymptotics of Cahn's equation, whereas the 1D radial growth of closed layers (spherical in the present model) occurs in a real grain structure; this gives a quite different long-time asymptotics.

The offered model seems to be closer to the actual grain-boundary network than Cahn's one and therefore should give a more realistic kinetics of transformation. In particular, the considered system directly goes to the actual grain structure after sintering. It should be noted that the first attempt to derive a VF equation for grain boundary nucleation was undertaken by Johnson and Mehl (Appendix D in ref. [4]) just for a matrix composed of spherical grains (of the same radius) under the assumption that a growing nucleus cannot cross grain boundaries. Both the Cahn and present models have no this constraint. Also, JM's VF equation drastically differs from the present one because of the quite different way of its derivation.

Earlier, the kinetics of grain-boundary nucleated phase transformations was investigated by computer simulations [34] and the Cahn model was tested. It was shown that the latter underestimates pronouncedly the transformation fraction in reality and an empirical equation with a fitting parameter was proposed to modify the Cahn model and reach agreement. Also, heterogeneous boundary nucleation was studied with the aid of $m$ -point correlation functions [35], this analysis was developed in subsequent works [36-39]. Spatiotemporal $m$ -point correlation functions were introduced by Secimoto [40]; in fact, the two-point one is employed for calculating the contribution of phantom nuclei to the VF in refs. [2, 7] as well as for calculating the variance of grains size distribution in refs. [41, 24].

The paper is organized as follows. In Section 2, the model is described and a nonisothermal VF equation is derived. The isothermal case is studied in detail in Section 3. The VF equation is obtained here in two variants, both as time- and radius-dependent. The Avrami exponent behavior is examined here as well. In Section 4, the case of instantaneous nucleation is considered. In Section 5, equations for the VF and Avrami exponent of an ensemble of size-distributed particles are derived and numerically examined for specific distribution functions. A model and VF equations for grain-boundary nucleated transformations are presented in Section 6. The case of diffusional growth is briefly considered in Section 7. Conclusions Section finalizes the paper.

## 2. Nonisothermal equations

We consider the process of crystallization of the spherical domain of volume $V_0 = (4\pi/3)R_0^3$ under the nucleation of new-phase centers on its surface with the nucleation rate $I_s(t)$ and growth velocity $u(t)$; both these quantities depend on temperature which varies with time, e. g., when the particle is heated or cooled with the constant rate $q$, $T = T_0 \pm qt$. Let us take at random the point $O'$ in the domain. Let it be at a distance $r$ from the centre of the domain (the point $O$), Fig. 1. We find the probability $Q(r,t)$ that the point $O'$ is untransformed at time $t$. For this purpose, we specify the *critical region* (CR) for the point $O'$ - the sphere of radius $R(t',t)$, Eq. (1a). At time $t'$, the boundary of this region moves at the velocity $u(t')$ towards the point $O'$, so that in the time interval $0 \leq t' \leq t$ its radius decreases from the greatest value $R(0,t) \equiv R_m(t)$ to $R(t,t) \equiv 0$. In order for the point $O'$ to be untransformed at time $t$, it is necessary and sufficient that no centre of a new phase should be formed within the CR in the time interval $0 \leq t' \leq t$. The probability of this event is [1, 11]

$$Q(r,t) = \exp[-Y(r,t)] \qquad (2)$$

In the case of bulk nucleation in infinite space, the function $Y$ does not depend on $r$ and has the following form [1]:

$$Y(t) = \int_0^t I_b(t')V(t',t)dt' \qquad (3)$$

where $V(t',t) = (4\pi/3)R^3(t',t)$ is the CR volume at time $t'$ (identical to the volume of a $t'$-age nucleus at time $t$). In the considered case, the new-phase centers appear on the domain surface only. We should take the part of surface lying within the CR (Fig. 1); its area is denoted by $\Omega(r;t',t)$. Applying the method of ref. [11], we replace $I_b(t')$ by $I_s(t')$ and $V(t',t)$ by $\Omega(r;t',t)$ in Eq. (3):

$$Y(r,t) = \int_0^t I_s(t')\Omega(r;t',t)dt' \qquad (4)$$

The VF $Q(t)$ of the material untransformed at time $t$ is the probability for the point $O'$ to fall into the untransformed part of the domain:

$$Q(t) = \frac{1}{V_0}\int_0^{R_0} Q(r,t)(4\pi r^2)dr \qquad (5)$$

Depending on $t$, $t'$, and $r$, the area $\Omega(r;t',t)$ can be equal to 0, $S(r;t',t)$, or $S_0 = 4\pi R_0^2$, where

$$S(r;t',t) = \frac{\pi R_0}{r}\left[R^2(t',t) - (R_0 - r)^2\right] \qquad (6)$$

Determining times $t_1$ and $t_2$ by the equations

$$R_m(t_1) = R_0, \quad R_m(t_2) = 2R_0 \qquad (7)$$

we have the following three cases with respect to time $t$, similarly to ref. [11].

(1) $R_m(t) < R_0$: $t < t_1$

We determine the distance $r_0 = R_0 - R_m$. At $0 \le r \le r_0$, the CR does not include the domain surface, accordingly, $\Omega(r; t', t) = 0$. We also determine the time $t_m(r,t)$ by the equation $R(t_m, t) = R_0 - r$. At $r_0 < r \le R_0$, the CR includes the part of the domain surface in the time interval $0 \le t' < t_m(r,t)$, accordingly, $\Omega(r; t', t) = S(r; t', t)$. Further, in the interval $t_m(r,t) \le t' \le t$ the CR is empty and $\Omega(r; t', t) = 0$. Summarizing,

$$Y_1(r,t) = \begin{cases} 0, & 0 \le r \le r_0 \\ \int\limits_0^{t_m(r,t)} I_s(t') S(r; t', t) dt', & r_0 < r \le R_0 \end{cases} \tag{8}$$

(2) $R_0 \le R_m(t) \le 2R_0$: $t_1 \le t \le t_2$.

We determine the distance $r_0' = R_m - R_0$ and the time $t_m'(r,t)$: $R(t_m', t) = R_0 + r$. Let us consider the case $0 < r < r_0'$. In Fig. 2, the CR boundary positions are shown for this case at different times $t'$. In the time interval $0 \le t' \le t_m'(r,t)$, the domain lies entirely within the CR, accordingly, $\Omega(r; t', t) = S_0$. In the interval $t_m'(r,t) < t' \le t_m(r,t)$, we have $\Omega(r; t', t) = S(r; t', t)$. And in the remaining interval $t_m(r,t) < t' \le t$, $\Omega(r; t', t) = 0$. At $r_0' \le r \le R_0$, we have $\Omega(r; t', t) = S(r; t', t)$ in the interval $0 \le t' < t_m(r,t)$ and $\Omega(r; t', t) = 0$ for $t_m(r,t) \le t' \le t$. In this way,

$$Y_2(r,t) = \begin{cases} S_0 \int\limits_0^{t_m'(r,t)} I_s(t') dt' + \int\limits_{t_m'(r,t)}^{t_m(r,t)} I_s(t') S(r; t', t) dt', & 0 \le r < r_0' \\ \int\limits_0^{t_m(r,t)} I_s(t') S(r; t', t) dt', & r_0' \le r \le R_0 \end{cases} \tag{9}$$

(3) $R_m(t) > 2R_0$: $t > t_2$

As it follows from the foregoing, in this case

$$\Omega(r; t', t) = \begin{cases} S_0, & 0 \le t' \le t_m'(r,t) \\ S(r; t', t), & t_m'(r,t) < t' < t_m(r,t) \\ 0, & t_m(r,t) \le t' \le t \end{cases} \tag{10}$$

for arbitrary $r$-value. Accordingly,

$$Y_3(r,t) = S_0 \int\limits_0^{t_m'(r,t)} I_s(t') dt' + \int\limits_{t_m'(r,t)}^{t_m(r,t)} I_s(t') S(r; t', t) dt', \quad 0 \le r \le R_0 \tag{11}$$

The VF $Q_i(t)$ in each case is calculated according to Eqs. (5) and (2); e.g., we have for case (1):

$$Q_1(t) = \frac{1}{V_0} \left\{ \int\limits_0^{r_0} (4\pi r^2) dr + \int\limits_{r_0}^{R_0} \exp\left[ -\int\limits_0^{t_m(r,t)} I_s(t') S(r; t', t) dt' \right] (4\pi r^2) dr \right\}$$

$$= \left(1 - \frac{R_m(t)}{R_0}\right)^3 + \frac{1}{V_0}\int\limits_{R_0-R_m}^{R_0} \exp\left[-\int\limits_0^{t_m(r,t)} I_s(t')S(r;t',t)dt'\right](4\pi r^2)dr \,, \quad t < t_1 \tag{12}$$

The VF at any time $t$ can be written with the aid of the Heaviside step function $\eta(x)$ as follows:

$$Q(t) = \eta(t_1 - t)Q_1(t) + \eta(t_2 - t)\eta(t - t_1)Q_2(t) + \eta(t - t_2)Q_3(t) \tag{13}$$

Accordingly, the transformed VF is $X(t) = 1 - Q(t)$.

## 3. Isothermal equations

### 3.1. Volume fraction as a function of time

The case of isothermal crystallization (constant $I_s$ and $u$) allows one to get an explicit time dependence of the VF and thereby to study the kinetics of crystallization in detail. The key quantities defined above acquire the following form:

$$R(t',t) = u(t-t') \,, \quad R_m(t) = ut \,, \quad r_0 = R_0 - ut \,, \quad t_m(r,t) = t - \frac{R_0-r}{u} \,, \quad r_0' = ut - R_0 \,,$$

$$t_m'(r,t) = t - \frac{R_0+r}{u} \tag{14}$$

We introduce the dimensionless time $\tau = ut/R_0$ and distance $x = r/R_0$ as well as the characteristic parameter $\alpha_s = (\pi/3)(I_s/u)R_0^3$. After the calculation of integrals in the above expressions, we get the following equations:

(1) $\tau < 1$:

$$Q_1(\tau) = (1-\tau)^3 + 3\int\limits_{1-\tau}^1 e^{-\alpha_s\varphi_1(x,\tau)} x^2 dx \,, \quad \varphi_1(x,\tau) = \frac{2(1-x)^3 - 3\tau(1-x)^2 + \tau^3}{x} \tag{15a}$$

(2) $1 \le \tau \le 2$:

$$Q_2(\tau) = 3\left\{\int\limits_0^{\tau-1} e^{-\alpha_s\varphi_2(x,\tau)} x^2 dx + \int\limits_{\tau-1}^1 e^{-\alpha_s\varphi_1(x,\tau)} x^2 dx\right\} \,, \quad \varphi_2(x,\tau) = 12\left(\tau - 1 - \frac{1}{3}x^2\right) \tag{15b}$$

(3) $\tau > 2$:

$$Q_3(\tau) = 3\int\limits_0^1 e^{-\alpha_s\varphi_2(x,\tau)} x^2 dx \tag{15c}$$

It should be noted that $\varphi_1(x,\tau) = 0$ at $x = 1 - \tau$ and $\varphi_1(x,\tau) = \tau^3$ at $x = 1$. These equations can be easily rewritten for the VF $X(\tau) = 1 - Q(\tau)$; e.g., Eq. (15a) goes into

$$X_1(\tau) = 3\int\limits_{1-\tau}^1 \left[1 - e^{-\alpha_s\varphi_1(x,\tau)}\right] x^2 dx \,, \quad \tau < 1 \tag{15d}$$

Isothermal equations for $X(t)$ were also obtained in ref. [20] by the causal-cone [42] or time-cone [43] method.

It is also of interest to get an equation for the untransformed fraction $Q_s(\tau)$ of the surface itself. It is given by Eq. (2) taken at $r = R_0$; according to Eqs. (8-11), the functions $\varphi_1(x,\tau)$ and $\varphi_2(x,\tau)$ are taken at $x = 1$. Doing so, we get

$$Q_s(\tau) = \begin{cases} e^{-\alpha_s \tau^3}, & 0 < \tau \le 2 \\ e^{-12\alpha_s(\tau - 4/3)}, & \tau > 2 \end{cases} \qquad (16)$$

It is seen that for $\tau < 2$ the law of spherical surface transformation is the same as the law of transformation of a plane, whereas the case of $\tau > 2$ shows the difference between them; it emphasizes the finiteness of geometry. So, the parameter $\alpha_s$ characterizes the rate of surface transformation; for sufficiently large values of $\alpha_s$ (this point is discussed in more detail below), the characteristic time of surface transformation is $\tau_s \sim \alpha_s^{-1/3}$.

In Fig. 3, the plots of $X(\tau)$ are shown for different $\alpha_s$ values together with the corresponding dependences for bulk nucleation from ref. [11]. The characteristic parameter for bulk nucleation is $\alpha_b = (\pi/3)(I_b/u)R_0^4$, where $I_b$ is the bulk nucleation rate; in order to compare both the dependences, we should take the corresponding value of $I_b$. The latter relates to the surface nucleation rate, $I_s$, as $I_b = \omega_0 I_s$, where $\omega_0$ is the surface area per unit volume. The number of nuclei emerging on the surface per unit time is $(4\pi R_0^2)I_s$; relating it to the particle volume, we get $\omega_0 = 3/R_0$ and $\alpha_b = 3\alpha_s$. The plots of $X_s(\tau) = 1 - Q_s(\tau)$ are also shown for comparison.

Transformation proceeds more rapidly for bulk nucleation, as it must; this effect is more pronounced for large $\alpha_s$s. The surface and particle transformations are close at small $\alpha_s$s, up to $\alpha_s = 1 \div 10$, and the former noticeably outstrips the latter at larger values; this effect is strongly pronounced for large $\alpha_s$s.

## 3.2. Avrami exponent

### 3.2.1. General remarks

As shown above, the AE naturally appears in *isothermal* transformations and reflects important features of the latter: the mechanisms of nucleation (homogeneous/heterogeneous) and growth (linear/diffusional), as well as space/growth dimensionality. Being a differential characteristic of a transformation process, it is more informative than a VF. Therefore, it is usually measured by experimenters as an important characteristic of a process. As follows from Eq. (1b),

$$n(t) = \frac{d\ln(-\ln Q(t))}{d\ln t} \qquad (17)$$

where the dependence $n(t)$ is written for generality; $n$ is constant for Eq. (1b) itself. It should be emphasized that though Eq. (17) was obtained within Kolmogorov's model, it is also applicable for more complex isothermal transformations, such as grain-boundary nucleated ones [33], transformations of spherical particles [11], etc; just in these cases $n$ varies with time, which reflects a change in the process.

On the other hand, the application of Eq. (18) to nonisothermal Eq. (1a) does not make sense; $n$ loses its informational content mentioned above due to the contribution from time-dependent nucleation and growth rates. At the same time, the mechanisms of nucleation and growth in a nonisothermal process are obviously the same as in isothermal one and a proper $n$ value for Eq. (1a) can be obtained as follows. In the case of heating with the rate $q$, $T = T_0 + qt$, we go from $t$ to $T$, $dt = dT / q$:

$$R(t',t) = R(T',T) = q^{-1}\int_{T'}^{T} u(\theta)d\theta, \quad V(T',T) = gR^D(T',T),$$

$$\int_{0}^{t} I(t')V(t',t)dt' = q^{-(D+1)}\int_{T_0}^{T} I(T')V(T',T)dT'$$

As a result,

$$X(T,q) = 1 - e^{-\chi(T)q^{-n}}, \quad \chi(T) \equiv \int_{T_0}^{T} I(T')V(T',T)dT' \qquad (18a)$$

$$\ln(-\ln Q(T,q)) = \ln \chi(T) - n\ln q, \quad n(q) = \frac{d\ln(-\ln Q(T,q))}{d\ln(1/q)} \qquad (18b)$$

where the dependence $n(q)$ is written for generality; $n$ is constant for Eq. (1a) itself. This corresponds to the familiar Ozawa equation [44].

At first glance, this method works only for transformations described by a simple exponential, as Eq. (1a); it is seen that the procedure of derivation of Eq. (18a) cannot be applied to Eq. (12). In order to understand how this method can be used for more complex transformations, we can consider the Cahn model [33] extended to a nonisothermal case in refs. [11, 31]. The VF equation in this case goes to Eq. (1a) at the early stage of transformation and to the equation

$$X(t) = 1 - e^{-2\omega y_m(t)} = 1 - e^{-2\omega q^{-1} y_m(T)} \quad , \qquad y_m(t) = \int_0^t u(\tau)d\tau = q^{-1} \int_{T_0}^{T(t)} u(\theta)d\theta \tag{19}$$

at the late stage (1D growth), so that the AE changes with time from 4 to 1. In isothermal case, this variation can be observed directly by measuring the VF as a function of time and using Eq. (18). In a nonisothermal case, the procedure is as follows.

A series of measurements of the VF $X_i(T, q_i)$, $i = 1, ..., k$, on $k$ identical samples during heating with the rates $q_i$ is performed. The curves $X_i(T, q_i)$ are arranged sequentially with increasing $q_i$. Further, we choose the temperature $T$ by vertical line. This line should intersect the first curve $X_1(T, q_1)$ near $X_1(T, q_1) = 1$ and the last curve $X_k(T, q_k)$ near $X_k(T, q_k) = 0$. Other curves are intersected somewhere between these values. These intersection points are plotted as $\ln(-\ln(1 - X_i(T, q_i)))$ vs. $\ln(1/q_i)$ and then connected by a *smooth* line; the slope tangent of this line gives $n(q)$ according to Eq. (19b). As intersection points correspond to different stages of transformation and the AE is different for these stages, the obtained dependence $n(q)$ should vary from 4 to 1. In order to get the reliable result, the number of points should be sufficient. It should be stressed that fitting the points by a straight line (as is often done in literature) will result here in loss of information about the grain-boundary nucleation mechanism. Straight-line fitting is applied only to transformations with constant $n$, i.e. in the case of Eq. (1a).

### 3.2.2. Avrami exponent for the present problem

So, we study the AE behavior $n(\tau) = d \ln(-\ln Q(\tau))/d \ln \tau$ for isothermal Eqs. (15a-c). Fig. 4 shows the dependence $n(\tau)$ for different values of the characteristic parameter $\alpha_s$. It is seen that this dependence is quite different for small and large values of the latter; it is smooth at a small $\alpha_s$ and exhibits "irregular" behavior at a large one. The only common features are the limit values: $n(\tau) \to 4$ at $\tau \to 0$ and $n(\tau) \to 1$ at $\tau \to \infty$.

The following expansion holds at $\tau \to 0$:

$$\int_{1-\tau}^{1} e^{-\alpha_s \varphi_1(x,\tau)} x^2 dx = \tau - \tau^2 + \frac{1}{3}\tau^3 - \frac{\alpha_s}{2}\tau^4 + ... \tag{20a}$$

Thus,

$$Q_1(\tau) = 1 - \frac{3}{2}\alpha_s \tau^4, \quad X_1(\tau) = \frac{3}{2}\alpha_s \tau^4 = \frac{1}{2}Y_b, \quad Y_b = 3\alpha_s \tau^4 = \frac{\pi}{3}I_b u^3 t^4, \quad I_b = \frac{3}{R_0}I_s \tag{20b}$$

This equation can be regarded as the first term of expansion of the exponential $Q_1(\tau) = \exp(-(1/2)Y_b(\tau))$; hence, $n = 4$ in this limit. $Y_b$ is the "extended volume" in the Avrami theory [8, 9]; the multiplier 1/2 reflects the fact that only half of the sphere born on the surface grows into the particle. So, despite the fact that nucleation occurs on the surface, the process of transformation qualitatively looks like the process of bulk nucleation within the spherical particle [11]: at the early stage, it is similar to random nucleation in infinite space with $n = 4$, which is described by Eqs. (1a, b). The same is true for the grain-boundary nucleated transformation in the Cahn model [33] mentioned above.

At the late stage, $\tau > 2$, we have

$$Q_3(\tau) = J(\alpha_s)e^{-12\alpha_s\tau} = J(\alpha_s)e^{-S_0 I_s t}, \quad J(\alpha_s) \equiv 3e^{12\alpha_s} \int_0^1 e^{4\alpha_s x^2} x^2 dx \qquad (21)$$

It is easy to get the law of approaching $n$ to unity:

$$n(\tau) \equiv \frac{1}{1 - z(\alpha_s)/\tau}, \quad z(\alpha_s) \equiv \frac{\ln J(\alpha_s)}{12\alpha_s} \qquad (22)$$

It should be stressed that $n = 1$ does not mean the 1D growth here, differently from Cahn's model; the term $\exp(-S_0 I_s t)$ in Eq. (21) is the probability that no center of new phase appears on the surface during time $t$ - this is approximately the probability for the particle to be untransformed at large times. The same asymptotics holds for the surface transformation, Eq. (16).

In this way, the AE generally varies with time from 4 to 1, similarly to the case of bulk nucleation [11]. However, the full range of this variation including the asymptotics $n \to 1$ can be observed only at small values of the parameter $\alpha_s$. In this case, the particle is gradually transformed by a small number of nuclei.

At large values of $\alpha_s$ (large $I$ /large $R_0$ /small $u$), the process of transformation proceeds quite differently. At the early stage, the surface is covered by the crust of new phase and further the *one-dimensional radial growth* (1DRG) of this crust occurs. Eq. (16) allows one to estimate the time $\tau_s$ of crust formation (Fig. 3b).

In the case of a large $\alpha_s$, the transformation is completed at $\tau \sim 1$ (the crust reaches the particle centre for this time), hence, we can restrict ourselves by the interval $\tau < 1$. Fig. 5 shows the contribution of each summand in Eq. (15a) to the VF $Q_1(\tau)$ in this case. It is seen that the contribution of the second (integral) summand is small and occurs at a short initial stage; after this stage,

$$Q_1(\tau) \approx (1-\tau)^3 = \frac{(R_0 - ut)^3}{R_0^3} \equiv Q^{(1D)}(\tau), \quad X^{(1D)}(\tau) = 1 - (1-\tau)^3 \tag{23}$$

which is the VFs in the case of 1DRG. The 1DRG curve in Fig. 3b is an asymptotic one for the VFs $X(\tau)$; the latter approach it at increasing $\alpha_s$ values.

The dependence $n(\tau)$ in Fig.4 exhibits the peak near $\tau = 1$; its origin is just the 1DRG. Indeed, the AE for Eq. (23) is

$$n^{(1D)}(\tau) = \tau \frac{d \ln(-\ln(1-\tau)^3)}{d\tau} = -\frac{\tau}{(1-\tau)\ln(1-\tau)} \tag{24}$$

This function is shown in Fig. 6; its limits are $n^{(1D)}(\tau) \to 1$ at $\tau \to 0$ and $n^{(1D)}(\tau) \to \infty$ at $\tau \to 1$. It is seen that the AE $n(\tau)$ at a large $\alpha_s$ abruptly drops from $n = 4$ to the dependence $n^{(1D)}(\tau)$ and further follows it. So, the dependence $n(\tau)$ for a large $\alpha_s$ in Fig. 4 is observable experimentally only for $\tau < 1$ (before the maximum).

In a nonisothermal case, the general picture of transformation is obviously the same; in particular, the 1DRG occurs as well. Hence, the AE behavior is the same, as in the isothermal case. Experimentally it can be revealed in the manner described above.

### 3.3. Volume fraction as a function of radius

Eqs. (15a-c) give the temporal dependence of the VF of a particle of radius $R_0$. It is also of interest to get the dependence of the VF on $R_0$ at a fixed time $t$. It is obtained from Eqs. (15a-c) by substitutions

$$\rho = \frac{1}{\tau} = \frac{R_0}{ut}, \quad y = \frac{r}{ut}, \quad x = \frac{y}{\rho} \tag{25}$$

The characteristic dimensionless parameter here is $\beta_s = (\pi/3)I_s u^2 t^3$; $\alpha_s = \beta_s \rho^3$. The result is as follows:

(1) $\rho \geq 1$:

$$Q_1(\rho) = \left(\frac{\rho - 1}{\rho}\right)^3 + \frac{3}{\rho^3} \int_{\rho-1}^{\rho} e^{-\beta_s \psi_1(y,\rho)} y^2 dy, \quad \psi_1(y,\rho) = \frac{\rho}{y}\left[2(\rho-y)^3 - 3(\rho-y)^2 + 1\right] \tag{26a}$$

(2) $\frac{1}{2} < \rho < 1$:

$$Q_2(\rho) = \frac{3}{\rho^3}\left\{ \int_0^{1-\rho} e^{-\beta_s \psi_2(y,\rho)} y^2 dy + \int_{1-\rho}^{\rho} e^{-\beta_s \psi_1(y,\rho)} y^2 dy \right\}, \quad \psi_2(y,\rho) = 12\left(\rho^2 - \rho^3 - \frac{1}{3}\rho y^2\right) \tag{26b}$$

(3) $0 < \rho \leq \dfrac{1}{2}$:

$$Q_3(\rho) = \frac{3}{\rho^3} \int_0^\rho e^{-\beta_s \psi_2(y, \rho)} y^2 dy \qquad (26c)$$

The VF of the transformed material is

$$X(\rho) = \eta\left(\frac{1}{2} - \rho\right) X_3(\rho) + \eta\left(\rho - \frac{1}{2}\right)\eta(1 - \rho) X_2(\rho) + \eta(\rho - 1) X_1(\rho) \qquad (27)$$

where $X_i(\rho) = 1 - Q_i(\rho)$. The dependence $X(\rho)$ is plotted in Fig. 7a. Differently from the case of bulk nucleation [11], where it is monotonically increasing, here the dependence $X(\rho)$ is not monotonic; it has a maximum at small $\beta_s$ values and the plateau $X(\rho) = 1$ for large $\beta_s$s. The number of nuclei is proportional to the surface area $S_0$. At the same time, the particle volume $V_0$ at small $R_0$ is sufficiently small, so that the increasing number of nuclei at increasing $R_0$ can transform the particle partially (at small times, or small $\beta_s$s) or entirely (at large $\beta_s$s). This explains the ascending branch of $X(\rho)$ and the plateau at small $R_0$. At large $R_0$, the volume effect prevails over the surface one, which results in the descending branch. The first summand in Eq. (26a) describes the 1DRG at large $\beta_s$s and $\rho > 1$.

The transformed VF $X_{pl}(t)$ for the case of nucleation on a plane was calculated in refs. [33, 11]; if we are interested in the one-sided problem only (filling space with growing hemispheres), the result is

$$X_{pl}(t) = \omega u t \int_0^1 \left[1 - e^{-\beta_s(1 - 3z^2 + 2z^3)}\right] dz \qquad (28a)$$

We can expect that this equation will approximate the exact solution, Eq. (26a), at large $R_0$, if $3/R_0$ is substituted for $\omega$:

$$X_{pl}(\rho) = \frac{3}{\rho} \int_0^1 \left[1 - e^{-\beta_s(1 - 3z^2 + 2z^3)}\right] dz \qquad (28b)$$

Indeed, Eq. (26a) for $X_1(\rho)$ is as follows:

$$X_1(\rho) = \frac{3}{\rho^3} \int_{\rho-1}^\rho \left[1 - e^{-\beta_s \psi_1(y, \rho)}\right] y^2 dy \qquad (29a)$$

Taking $z = \rho - y$ as a new integration variable, $dy = -dz$, we obtain the following equation:

$$X_1(\rho) = \frac{3}{\rho} \int_0^1 \left[1 - e^{-\beta_s \psi_1(z, \rho)}\right]\left(1 - \frac{z}{\rho}\right)^2 dz, \quad \psi_1(z, \rho) = \frac{1}{1 - z/\rho}\left[1 - 3z^2 + 2z^3\right] \qquad (29b)$$

It is seen that for $\rho \gg 1$ and therefore $z/\rho \ll 1$, $X_1(\rho)$ is close to $X_{pl}(\rho)$. Fig. 7a shows how Eq. (28b) approaches the exact solution; the approximation begins with smaller radii, when $\beta_s$ decreases.

Similarly to the VF, the AE also can be represented as a function of the particle radius, $n(\rho)$. Indeed, as the function $n(\tau)$ depends on the parameter $\alpha_s = \beta_s \rho^3$ as well, we have $n(\rho) = n(\alpha_s(\rho), \tau(\rho)) = n(\beta_s \rho^3, 1/\rho)$. The dependence $n(\rho)$ is shown in Fig. 7b for different $\beta_s$s. The view is mirror with respect to Fig. 4, except the fact that at $\rho \to \infty$ the AE does not approach the value $n = 4$; this is due to the fact that $\alpha_s$ increases together with $\rho$. The AE is close to 4 at small $\beta_s$s and approach the unity at large ones.

## 4. Instantaneous nucleation

Differently from continuous nucleation considered above, here the nuclei appear at $t' = 0$. If $N_s$ is the mean surface density of these centers, then the nucleation rate can be represented with the aid of the $\delta$-function as $I_s(t') = N_s \delta_+(t')$. As before, we use the dimensionless variables $\tau$ and $x$ as well as the characteristic parameter $\gamma_s = N_s S_0 = 4\pi N_s R_0^2$ which is the mean number of nuclei on the surface (the meaning of this phrase is explained later). Substituting $R(t', t) = R(0, t) = R_m(t) = ut$ in Eq. (6), we get $S(r; 0, t) = S_0 \varphi_0(x, \tau)$, where

$$\varphi_0(x, \tau) = \frac{\tau^2 - (1-x)^2}{4x} \tag{30}$$

The VF equations are obtained from general equations of Section II as follows:

(1) $\tau < 1$:

$$Q_1(\tau) = (1-\tau)^3 + 3 \int_{1-\tau}^{1} e^{-\gamma_s \varphi_0(x, \tau)} x^2 dx \tag{31a}$$

(2) $1 \le \tau < 2$:

$$Q_2(\tau) = (\tau-1)^3 e^{-\gamma_s} + 3 \int_{\tau-1}^{1} e^{-\gamma_s \varphi_0(x, \tau)} x^2 dx \tag{31b}$$

(3) $\tau \ge 2$:

$$Q_3(\tau) = e^{-\gamma_s} \tag{31c}$$

The surface itself is transformed according to the equations

$$Q_s(\tau) = \begin{cases} e^{-\gamma_s \tau^2/4}, & 0 < \tau < 2 \\ e^{-\gamma_s}, & \tau \ge 2 \end{cases} \tag{32}$$

The dependences $X(\tau) = 1 - Q(\tau)$ are shown in Fig. 8 for different $\gamma_s$'s. Differently from the case of continuous nucleation, the VF $X(\tau)$ does not reach unity now; its final value is $1 - \exp(-\gamma_s)$. Indeed, the maximal time of particle transformation here is $\tau = 2$. It occurs, if *one* nucleus appears at $t' = 0$; for this time, it goes the path $2R_0$ and entirely covers the particle. If this event does not occur (no nucleus appears), the particle remains untransformed; Eq. (31c) gives the probability of this case. The distinction of the final VF from unity is observable only at small $\gamma_s$ s. So, the time of transformation in the case of instantaneous nucleation is *finite* and does not exceed $\tau = 2$. For large $\gamma_s$ s, the main features of transformation are the same as in Fig. (3b) for continuous nucleation.

The AE varies with time here from 3 to 0. The dependence $n(\tau)$ is smooth at small $\gamma_s$'s and "irregular" at large ones due to the 1DRG, similarly to the case of continuous nucleation. For large $\gamma_s$ s, the AE behavior is similar to that of Fig. 6: $n(\tau)$ drops from $n = 3$ to the 1DRG curve and further follows it.

The variables of Eq. (25) and the characteristic parameter $\lambda_s = 4\pi N_s u^2 t^2 = \gamma_s \tau^2$, $\gamma_s = \lambda_s \rho^2$, are employed to get a radius-dependent VF:

(1) $\rho \geq 1$:

$$Q_1(\rho) = \left(\frac{\rho - 1}{\rho}\right)^3 + \frac{3}{\rho^3} \int_{\rho-1}^{\rho} e^{-\lambda_s \psi_0(y,\rho)} y^2 dy, \quad \psi_0(y,\rho) = \frac{\rho}{4y}\left[1 - (\rho - y)^2\right] \qquad (33a)$$

(2) $\frac{1}{2} < \rho < 1$:

$$Q_2(\rho) = \left(\frac{1 - \rho}{\rho}\right)^3 e^{-\lambda_s \rho^2} + \frac{3}{\rho^3} \int_{1-\rho}^{\rho} e^{-\lambda_s \psi_0(y,\rho)} y^2 dy \qquad (33b)$$

(3) $0 < \rho \leq \frac{1}{2}$:

$$Q_3(\rho) = e^{-\lambda_s \rho^2} \qquad (33c)$$

The dependence $X(\rho)$ for different $\lambda_s$'s exhibits the behavior similar to that for continuous nucleation.

## 5. The volume fraction of an ensemble of particles

The above VF equations give the VF value $X(\tau)$ averaged over an ensemble of identical particles. The VF value $X_i(\tau)$ for a specific particle is obviously a random quantity as a result of realization of the random process. The number $N_{en}$ of particles in the ensemble should be sufficiently large; it depends on the variance $\sigma_X(\tau) = \overline{[X_i(\tau) - X(\tau)]^2}$. The evaluation of this variance is a separate statistical problem which is beyond the scope of the present paper. A brief analysis of this issue in ref. [11] and estimate on the basis of central limit theorem give $N_{en} \approx \max[(3\sigma_X(\tau)/\varepsilon)^2]$, where $\varepsilon$ is the required accuracy – the deviation of the ensemble-averaged VF value from $X(\tau)$.

The discussed issue is especially clear in the case of instantaneous nucleation. The number of nuclei appearing on the surface of a specific particle is a random quantity including zero; $\gamma_s = N_s S_0$ is its mean (the parameter of Poisson distribution). Accordingly, the VF's $X_i(\tau)$ in different particles are different; even the particles with the same numbers of nuclei have different VF's due to different locations of these nuclei. Eqs. (31a-c) give the ensemble-averaged VF value; this averaging is double: first, the averaging over all positions of nuclei for the particles with the same number of the latter and second, the averaging over all numbers of nuclei with the Poisson distribution [11]. The necessity to solve this statistical problem *ab initio* is avoided by the use of the present method. Eq. (31c) can be also treated as the relative number of particles without the nucleation event.

So, all the above VF equations can be applied in practice to ensembles of a large number of identical particles or particles with close sizes. If the sizes of particles in an ensemble are different, we need to derive a VF equation for this ensemble. Just the radius-dependent VF's $X(\kappa, \rho)$ derived above are suitable for this purpose, where $\kappa$ is the characteristic parameter, $\kappa = \beta_s$ or $\lambda_s$.

Let $f(R_0)$ be the size distribution function of particles; $f(R_0)dR_0$ is the number of particles with sizes within the interval $[R_0, R_0 + dR_0]$. We assume that this number is large, so that Eqs. (26a-c) or (33a-c) give the untransformed VF of this subsystem. Then the transformed VF of an ensemble is

$$X_{en}(\eta) = \frac{\int\limits_{R_0^{(1)}}^{R_0^{(2)}} X(\kappa, \rho) V_0 f(R_0) dR_0}{\int\limits_{R_0^{(1)}}^{R_0^{(2)}} V_0 f(R_0) dR_0} \tag{34}$$

where the transformed and full volumes of particles in the ensemble are in the numerator and denominator, respectively; $R_0^{(1)}$ and $R_0^{(2)}$ are the smallest and the largest sizes in the ensemble.

To get the final expression, the following relations are utilized:

$$\beta_s = b(ut)^3, \ b \equiv \frac{\pi}{3}\frac{I_s}{u}, \ ut = \left(\frac{\beta_s}{b}\right)^{1/3}; \ \lambda_s = a(ut)^2, \ a \equiv 4\pi N_s, \ ut = \sqrt{\frac{\lambda_s}{a}} \quad (35)$$

Thus, $\rho(\beta_s) = R_0 / ut = R_0 / \sqrt[3]{\beta_s / b}$ for continuous nucleation and $\rho(\lambda_s) = R_0 / \sqrt{\lambda_s / a}$ for instantaneous one; $a$ and $b$ are the input parameters. Accordingly, VF equations for both the cases are

$$X_{en}(\beta_s) = \frac{\int_{R_0^{(1)}}^{R_0^{(2)}} X\left(\beta_s, \frac{R_0}{\sqrt[3]{\beta_s / b}}\right) R_0^3 f(R_0) dR_0}{\int_{R_0^{(1)}}^{R_0^{(2)}} R_0^3 f(R_0) dR_0}, \quad X_{en}(\lambda_s) = \frac{\int_{R_0^{(1)}}^{R_0^{(2)}} X\left(\lambda_s, \frac{R_0}{\sqrt{\lambda_s / a}}\right) R_0^3 f(R_0) dR_0}{\int_{R_0^{(1)}}^{R_0^{(2)}} R_0^3 f(R_0) dR_0} \quad (36)$$

It should be noted in this connection that the AE $n(\rho)$, Fig. 7b, can be transformed to the dependence $n(R_0)$ with the aid of Eq. (35) as well:

$n(\beta_s, R_0) = n(\alpha_s(R_0), 1/\rho(\beta_s)) = n(bR_0^3, \sqrt[3]{\beta_s / b} / R_0)$.

Employing the mean size $\overline{R}_0$ of the particle and the relations

$$\alpha_s = \frac{\pi}{3}\frac{I_s}{u}\overline{R}_0^3 = b\overline{R}_0^3, \quad \gamma_s = 4\pi N_s \overline{R}_0^2 = a\overline{R}_0^2, \quad \tau = \frac{ut}{\overline{R}_0} \quad (37a)$$

we get

$$\beta_s = \alpha_s \tau^3, \quad \lambda_s = \gamma_s \tau^2 \quad (37b)$$

With the aid of these relations, the VF's $X_{en}$ are represented as functions of time:

$X_{en}(\beta_s) = X_{en}(\alpha_s \tau^3)$ and $X_{en}(\lambda_s) = X_{en}(\gamma_s \tau^2)$.

Calculating the derivative of double logarithm with the use of relation $d/d\tau = (d\kappa/d\tau)(d/d\kappa)$, we get AEs for an ensemble in both the cases:

$$n(\beta_s) = \frac{3\beta_s}{Q_{en}(\beta_s)\ln Q_{en}(\beta_s)}\frac{dQ_{en}(\beta_s)}{d\beta_s}, \quad n(\lambda_s) = \frac{2\lambda_s}{Q_{en}(\lambda_s)\ln Q_{en}(\lambda_s)}\frac{dQ_{en}(\lambda_s)}{d\lambda_s} \quad (38)$$

where $Q_{en} = 1 - X_{en}$ and the AE temporal evolution is found with the aid of Eq. (37b) again:

$n(\beta_s) = n(\alpha_s \tau^3)$ and $n(\lambda_s) = n(\gamma_s \tau^2)$.

Amorphous powders studied by experimentalists are usually obtained as a result of grinding. The size distribution function (DF) of particles for this process was obtained by Kolmogorov [45] under certain model assumptions; this is the logarithmic normal distribution:

$$f(r) = \frac{1}{r\sigma\sqrt{2\pi}}e^{-\frac{(\ln r - \mu)^2}{2\sigma^2}} \quad (39a)$$

In subsequent studies of this issue [46-48], the process of grinding was considered in the framework of physical approaches and the validity of Kolmogorov's distribution was confirmed for a certain physical model as well [47]; at the same time, the power asymptotic forms $f(r) \sim r^{-5}$ [46] and $f(r) \sim r^{-6}$ [48] at $r \to \infty$ were obtained for other models.

The mean and variance for this DF are respectively

$$\bar{r} = e^{\mu + \sigma^2/2}, \ \sigma_r^2 = \overline{(r - \bar{r})^2} = \left(e^{\sigma^2} - 1\right)e^{2\mu + \sigma^2}$$

from where

$$\sigma^2 = \ln\left(1 + \frac{\sigma_r^2}{\bar{r}^2}\right), \ \ \mu = \ln\bar{r} - \frac{1}{2}\ln\left(1 + \frac{\sigma_r^2}{\bar{r}^2}\right)$$

The dimensionless radius $r = R_0 / \bar{R}_0$ is employed here, hence, $\bar{r} = 1$ and

$$\sigma^2 = \ln\left(1 + \sigma_r^2\right), \quad \mu = -(1/2)\sigma^2 \tag{39b}$$

It is seen from Eq. (37a) that a change in $\bar{R}_0$ results in the change of $\alpha_s$ value. The VF and AE behavior at different $\alpha_s$s was considered above. Therefore, it is necessary to examine the behavior of these quantities for different rms values $\sigma_r$ (the distribution width) only. Some range of $\sigma_r$ values is used for this purpose. In experimental conditions, the $\sigma_r$ value can be found from fitting the obtained distribution by Eq. (39a). The case of instantaneous nucleation is used here for numerical examples.

As follows from the above relations, $R_0 = r\bar{R}_0$, $\bar{R}_0\sqrt{a} = \sqrt{\gamma_s}$, $\sqrt{\gamma_s / \lambda_s} = 1/\tau$, so that Eq. (36) acquires the form

$$X_{en}(\tau) = \frac{\int_0^\infty X\left(\gamma_s \tau^3, \frac{r}{\tau}\right) r^3 f(r) dr}{\int_0^\infty r f(r) dr} \tag{39c}$$

i.e. the VF temporal dependence is governed by the parameter $\gamma_s$ only, as for the ensemble of identical particles of radius $\bar{R}_0$.

Another DF employed here is the uniform one,

$$f(R_0) = \frac{1}{R_0^{(2)} - R_0^{(1)}} \tag{40a}$$

Eq. (36) for this DF acquires the form

$$X_{en}(\lambda_s) = \frac{\int_{R_0^{(1)}}^{R_0^{(2)}} X\left(\lambda_s, \frac{R_0}{\sqrt{\lambda_s / a}}\right) R_0^3 dR_0}{\left((R_0^{(2)})^4 - (R_0^{(1)})^4\right)/4} \tag{40b}$$

As shown above, the results do not depend on length units; $\bar{R}_0 = 200$ of conventional units, say, $\mu m$, was put for the uniform DF. The input parameter $a$ is chosen to get a desired $\gamma_s$ value, according to Eq. (37a). Also, it was put $R_0^{(1)} = 0$, $R_0^{(2)} = 400$; this case is also called here as wide uniform distribution. As another variant, a narrow uniform DF with $R_0^{(1)} = 170$ and $R_0^{(2)} = 230$ is considered.

The results of numerical calculations are shown in Fig. 9. Fig. (a) shows that the maximal VF value increases with increasing distribution width $\sigma_r$ evidently due to the contribution from large particles on the tail of distribution ($\gamma_s \sim R_0^2$). For large $\sigma_r$ s, the $X(\tau)$ curve is strongly "stretched" and shifted in time due to the long tail of this DF again. The AE in Fig. (b) varies with time from 3 to 0, as mentioned above; the size distribution "stretches" this variation in time. Fig. (c) shows how the transformation kinetics slows down with successive increase in $\sigma_r$ at a large $\gamma_s$ value.

Finally, Fig. (d) shows the AE behavior at a large $\gamma_s$, where the 1DRG occurs. For the ensemble of identical spheres ($\delta$), it is similar to that of Fig. 6: the AE drops from $n = 3$ to the 1DRG curve, Eq. (24), and further follows it. As follows from Fig. 7a, different particles in an ensemble with size distribution have different degree of transformation at a given time; the 1DRG occurs in large particles. Therefore, the distribution deflects the AE curve from the 1DRG one. For small $\sigma_r$ s, this effect is small, as it must; it increases with increase in $\sigma_r$. As is seen, the AE-curve slope even becomes negative at sufficiently large $\sigma_r$, so that the curve becomes monotonically decreasing. This is a specific feature of the logarithmic normal DF; for more "compact" DFs, e.g., the normal one, the AE-curve slope remains positive up to the largest $\sigma_r$ values.

It is interesting also to see the AE behavior for the uniform DFs. For the narrow one, it is close to that for the $\delta$-distribution, as it was expected. For the wide one, the AE curve is similar to that for the $\delta$-distribution and goes "in parallel" to it at some distance.

## 6. Model for grain-boundary nucleated transformations

### 6.1. Nonisothermal equations

We begin with the system of identical particles (grains), as before, to get the time-dependent VF. Taking one of the grains (Fig.10) and calculating the VF $Q(t)$ for it, we should take into account the possibility for the point $O'$ to be transformed by a nucleus appeared on boundaries of other grains as well; i.e. boundaries within the CR segment $ABC$ outside the given grain. The volume of this segment is (cf. ref. [11])

$$v(r_1, r_2; h) = \pi \left\{ \frac{2}{3}(r_2^3 - r_1^3) - \frac{h^3}{12} + \frac{1}{2}h(r_1^2 + r_2^2) + \frac{1}{4}\frac{(r_1^2 - r_2^2)^2}{h} \right\} \tag{41}$$

with $r_2 = R(t', t)$, $r_1 = R_0$, and $h = r$, so that $v = v(r; t', t)$. This is the possibility for a growing nucleus to cross grain boundaries which is forbidden by JM's model.

As a result, the functions $Y_i(r, t)$ of Section II denoted below as $Y_i^{(s)}(r, t)$ acquire one more summand $Y_i^{(b)}(r, t)$ which is the contribution from external boundaries. For the area $S^{(b)}(r; t', t)$ of these boundaries within the segment $ABC$, we employ the mean value

$$S^{(b)}(r; t', t) = \omega v(r; t', t) \tag{42}$$

where $\omega$ is the area of grain boundaries in unit volume.

In view of the relation $\omega I_s = I_b$, equations for $Y_i^{(b)}(r, t)$ are as follows:

$$Y_1^{(b)}(r, t) = \begin{cases} 0, & 0 \le r \le r_0 \\ \int_0^{t_m(r,t)} I_b(t') v(r; t', t) dt', & r_0 < r \le R_0 \end{cases} \tag{43a}$$

$$Y_2^{(b)}(r, t) = \begin{cases} \int_0^{t'_m(r,t)} I_b(t')[V(t', t) - V_0] dt' + \int_{t'_m(r,t)}^{t_m(r,t)} I_b(t') v(r; t', t) dt', & 0 \le r < r'_0 \\ \int_0^{t_m(r,t)} I_b(t') v(r; t', t) dt', & r'_0 \le r \le R_0 \end{cases} \tag{43b}$$

$$Y_3^{(b)}(r, t) = \int_0^{t'_m(r,t)} I_b(t')[V(t', t) - V_0] dt' + \int_{t'_m(r,t)}^{t_m(r,t)} I_b(t') v(r; t', t) dt', \quad 0 \le r < R_0 \tag{43c}$$

In other words, the influence of external boundaries is represented as some effective bulk nucleation outside the given grain. Finally,

$$Y_i(r, t) = Y_i^{(s)}(r, t) + Y_i^{(b)}(r, t) \tag{44}$$

### 6.2. Isothermal kinetics

Calculating the integrals in Eqs. (43a-c) at constant $I_b$, $u$ as well as using Eqs. (2) and (5), we get the characteristic parameter $\alpha_b = (\pi/3)(I_b/u)R_0^4 = c\alpha_s$, $c \equiv \omega R_0$; in view of this relation, the final VF equations have the following form:

(1) $\tau < 1$:

$$Q_1^{(gb)}(\tau) = (1-\tau)^3 + 3\int_{1-\tau}^{1} e^{-\alpha_s[\varphi_1(x,\tau) + c\varphi_1^{(b)}(x,\tau)]} x^2 dx \tag{45a}$$

(2) $1 \leq \tau \leq 2$:

$$Q_2^{(gb)}(\tau) = 3\left\{ \int_0^{\tau-1} e^{-\alpha_s[\varphi_2(x,\tau) + c\varphi_2^{(b)}(x,\tau)]} x^2 dx + \int_{\tau-1}^{1} e^{-\alpha_s[\varphi_1(x,\tau) + c\varphi_1^{(b)}(x,\tau)]} x^2 dx \right\} \tag{45b}$$

(3) $\tau > 2$:

$$Q_3^{(gb)}(\tau) = 3\int_0^{1} e^{-\alpha_s[\varphi_2(x,\tau) + c\varphi_2^{(b)}(x,\tau)]} x^2 dx \tag{45c}$$

where the functions $\varphi_i(x,\tau)$ are given by Eqs. (15a, b), as before, whereas the functions $\varphi_i^{(b)}(x,\tau)$ represent the contribution from external boundaries:

$$\varphi_1^{(b)}(x,\tau) = \frac{3}{20}\frac{[\tau^5 - (1-x)^5]}{x} + \frac{1}{2}[\tau^4 - (1-x)^4] + \frac{1}{2}\left(x - \frac{1}{x}\right)[\tau^3 - (1-x)^3]$$

$$+ \left(\frac{3}{4x} - 2 + \frac{3}{2}x - \frac{1}{4}x^3\right)[\tau - (1-x)] \tag{46a}$$

$$\varphi_2^{(b)}(x,\tau) = \tau^4 - 4\tau - \frac{1}{5}x^4 + 2x^2 + 3 \tag{46b}$$

For numerical calculations, we should have some estimate for $\omega$. Considering a large number of spheres packed into the cubic lattice, $n$ spheres per edge, we have approximately $(2R_0 n)^3$ for the volume of this system; the total area of boundaries is $4\pi R_0^2 n^3$. Hence, $\omega = (\pi/2)(1/R_0)$ and $c = \pi/2 = 1.57$ for this system. This value will slightly vary for other types of packing including the random dense one. Cahn [33] used a certain structural model to evaluate $\omega$. Assuming that all grains are represented by equal truncated octahedrons (polyhedra whose faces are squares and hexagons) and $D$ is the distance between square faces, a value $\tilde{c} = 3.35$ was obtained for the relation $\omega = \tilde{c}/D$. As $D$ corresponds to $2R_0$, one obtains $c = \tilde{c}/2 = 1.67$. Differently from spheres, these polyhedra fill space without emptiness. On the other hand, the quantity $\omega$ can be employed as an adjustable parameter here.

In order to convert Eqs. (45a-c) to radius-dependent VF's, we use Eq. (25) and the characteristic parameter $\beta_b = (\pi/3)I_b u^3 t^4$, $\alpha_b = \beta_b \rho^4$. In view of Eq. (35),

$$\beta_b = \beta_s \omega u t = \beta_s \omega (\beta_s/b)^{1/3} \tag{47}$$

As a result,

(1) $\rho \geq 1$:

$$Q_1^{(gb)}(\rho) = \left(\frac{\rho-1}{\rho}\right)^3 + \frac{3}{\rho^3} \int_{\rho-1}^{\rho} e^{-\beta_s [\psi_1(y,\rho) + \omega(\beta_s/b)^{1/3} \psi_1^{(b)}(y,\rho)]} \, y^2 dy \qquad (48a)$$

(2) $\frac{1}{2} < \rho < 1$:

$$Q_2^{(gb)}(\rho) = \frac{3}{\rho^3} \left\{ \int_0^{1-\rho} e^{-\beta_s [\psi_2(y,\rho) + \omega(\beta_s/b)^{1/3} \psi_2^{(b)}(y,\rho)]} \, y^2 dy + \int_{1-\rho}^{\rho} e^{-\beta_s [\psi_1(y,\rho) + \omega(\beta_s/b)^{1/3} \psi_1^{(b)}(y,\rho)]} \, y^2 dy \right\} \qquad (48b)$$

(3) $0 < \rho \le \frac{1}{2}$:

$$Q_3^{(gb)}(\rho) = \frac{3}{\rho^3} \int_0^{\rho} e^{-\beta_s [\psi_2(y,\rho) + \omega(\beta_s/b)^{1/3} \psi_2^{(b)}(y,\rho)]} \, y^2 dy \qquad (48c)$$

where the functions $\psi_i(y,\rho)$ are given by Eqs. (26a, b), whereas the functions $\psi_i^{(b)}(y,\rho)$ are as follows:

$$\psi_1^{(b)}(y,\rho) = \frac{3}{20y}[1-(\rho-y)^5] + \frac{1}{2}[1-(\rho-y)^4] + \frac{1}{2}\left(y - \frac{\rho^2}{y}\right)[1-(\rho-y)^3]$$

$$+ \left(\frac{3}{4}\frac{\rho^4}{y} - 2\rho^3 + \frac{3}{2}y\rho^2 - \frac{1}{4}y^3\right)[1-(\rho-y)] \qquad (49a)$$

$$\psi_2^{(b)}(y,\rho) = 1 - 4\rho^3 + 3\rho^4 - (1/5)y^4 + 2y^2\rho^2 \qquad (49b)$$

Further, the transformed VF $X^{(gb)}(\beta_s,\rho) = 1 - Q^{(gb)}(\beta_s,\rho)$ of the $\rho$ - fraction is substituted in Eq. (36) for continuous nucleation to get the transformed VF for grain boundary nucleation. Thereafter, the mean grain size $\overline{R}_0$ is employed as a characteristic length and the characteristic parameter $\alpha_s$ and the time $\tau$ are expressed in terms of $\overline{R}_0$ according to Eq. (37a); $\beta_s$ relates to $\alpha_s$ and $\tau$ according to Eq. (37b). Finally,

$$X_{gb}(\tau) = \frac{\int_0^{\infty} X^{(gb)}\left(\alpha_s \tau^3, \frac{r}{\tau}\right) r^3 f(r) dr}{\int_0^{\infty} r^3 f(r) dr} \qquad (50)$$

The obtained VF is governed by the parameter $\alpha_s$ only, as for the system of identical spheres of radius $\overline{R}_0$.

It is worth noting that if the actual grain structure is a result of the nucleation-and-growth process with the known nucleation and growth rates, then the quantity $\omega$ can be calculated exactly for this structure; the following equation was derived in ref. [49]:

$$\omega(t) = \frac{2}{3} \int_0^t \frac{(dQ(\tau)/d\tau)^2}{u(\tau)Q(\tau)} d\tau \qquad (51)$$

In the isothermal case,

$$Q(\tau) = \begin{cases} \exp(-k\tau^4), \ k = (\pi/3)I_b u^3 \\ \exp(-k\tau^3), \ k = (4\pi/3)N_v u^3 \end{cases} \qquad (52)$$

are substituted in Eq. (51) for continuous and instantaneous nucleation. The mean grain volume $\overline{V_0}$ in the final state is given by Eq. (A13) of Appendix for continuous nucleation and $\overline{V_0} = N_v^{-1}$ for instantaneous one; $\overline{R_0} = (3/4\pi)^{1/3} \overline{V_0}^{1/3}$. Finally, one obtains

$$c = \omega \overline{R_0} = \begin{cases} 2\Gamma(3/4)[\Gamma(1/4)]^{-1/3} = 1.6, & \text{continuous} \\ (4/3)\Gamma(2/3) = 1.8, & \text{instantaneous} \end{cases} \qquad (53)$$

The VF equation by Cahn [33] in our notations is

$$X_C(\tau) = 1 - e^{-X_e(\tau)}, \quad X_e(\tau) = 2c\tau \int_0^1 \left[1 - e^{-\alpha_s \tau^3 (1-3x^2+2x^3)}\right] dx \qquad (54)$$

The KJMA equation for bulk nucleation is

$$X_K(\tau) = 1 - \exp(-c\alpha_s \tau^4) \qquad (55)$$

If the grain structure was obtained as a result of instantaneous nucleation, then the $\Gamma$-distribution

$$f(\upsilon) = \frac{p(p\upsilon)^{p-1}}{\Gamma(p)} e^{-p\upsilon} , \quad \upsilon = \frac{V}{\overline{V_0}} \qquad (56)$$

is assumed to be well established for it and $p = 5.586$ was obtained [24]. However, there is no closed expression for the case of continuous nucleation, where the DF is obtained as a result of solving the Fokker-Planck-like equation and computer simulations [22-26]. Johnson and Mehl [4] proposed a simple but effective idea to approximate the actual DF by the so-called "age-distribution". First, this approach allows one to derive a closed DF expression even in a nonisothermal case; second, it can be extended to a more realistic distribution. These steps are performed in Appendix.

The extended JM's DF, Eq. (A17), is used here for numerical calculations. Aiming the qualitative results only, a constant value of $\xi_x$ in Eq. (A16), $\xi_x = 0.2$, is employed. The plots for volume and radius DFs are shown in Fig. 11. While the original JM's DF ends at the maximal size, the extended one is continuous due to the fact that now there are large grains in the system with vanishingly small probability.

### 6.3. Discussion of the present model

In Fig. 12, VFs and AEs for the system of identical spheres, Eqs. (45a-c) are presented; $c = 1.6$ was put. Among the three VF curves – the present, Cahn, and KJMA – the Cahn ones are most sensitive to $c$ values due to the 1D asymptotics $X_C(\tau) = 1 - \exp[-2c\tau]$ at the late stage. At the same time, the AEs for Cahn's and KJMA's models do not depend on $c$, since the VF equations has the form $X_C(\tau) = 1 - \exp[-cg(\tau)]$. Fig. (12a) shows that the Cahn model underestimates VFs in comparison with the present and KJMA ones, which agrees with previous studies [34]. Both the present and Cahn curves have a common part at the initial stage; the size of this part increases with increasing $\alpha_s$. E.g., at $\alpha_s = 10^4$ both the curves begin to diverge at $X \approx 0.5$.

At a small $\alpha_s$, the present and KJMA curves are close to each other, as it was expected. Small $\alpha_s$ means small $R_0$ or/and $I_s$; in the first case, the system becomes more "homogeneous", which diminishes the correlations in the location of nuclei. In this way, the transformation process in the present model proceeds as in the KJMA one at random nucleation; the Cahn model is close to these two at the early stage and deviates at the late one due to the 1D asymptotics which is inherent in this model independently of parameters. These conclusions are supported by the AE behavior in Fig. (12b). AE in the present model is close to the value $n = 4$ of the KJMA one at all times; the corresponding $\ln(-\ln Q(\tau))$ -vs.- $\ln \tau$ plot is a straight line indistinguishable from the KJMA one. At the same time, the Cahn AE changes from 4 to 1.

So, the influence of external boundaries in the present model (the possibility for a growing nucleus to cross grain boundaries forbidden by JM's model) is important at small $\alpha_s$ values. Just due to this effect, we have the mentioned similarity between the present and KJMA models. In the JM's model, we would have the AE variation between 4 and 1 inherent for a single particle (Fig. (4)). Also, the comparison of Figs. (12a) and (3a) for $\alpha_s = 0.1$ shows that the influence of external boundaries approximately twice accelerates the transformation process.

At a large $\alpha_s$, the influence of external boundaries is negligible due to the transformation of grain boundary at the early stage with subsequent 1DRG (a nucleus from the outside cannot penetrate into the given grain). Thus the possibility for a growing nucleus to cross grain boundaries is inessential in this case and the transformation process is similar to that for a single particle. Accordingly, the AE plot is the same as in Fig. 6. As before, the Cahn AE changes from 4 to 1, but already at small times because of the large $\alpha_s$. Both the present and Cahn curves have a common part at the initial stage and drastically diverge at the late one. Also, the VF curve of the present model, Fig. (12a), is similar to that in Fig. 3b; it follows the 1DRG asymptotics.

Differing from each other, both the present and Cahn VFs strongly differ from the KJMA one at a large $\alpha_s$; this is due to quite different patterns of space filling in all three models.

Fig. 13 shows how the above size distributions change the plots of Fig. 12; the $c$ values by Eq. (53) were employed – 1.8 for the gamma DF and 1.6 for the extended JM one. While the Cahn VFs underestimate the present ones at a small $\alpha_s$, as before, the situation is somewhat different at a large one; here the Cahn curves cross the present ones and underestimate them only at the late stage, where transition to the 1D asymptotics takes place. This occurs due to the fact that the size distributions deflect the present-model curves from the 1DRG one for identical spheres, whereas the Cahn curves are insensitive to size distributions. The latter are governed by the parameter $c$ which is not connected directly with size distributions in the present approach; there is only the correspondence between them, as described above. In this way, the Cahn curve with $c = 1.6$ stays in place, whereas the corresponding "ex_JM" curve is shifted noticeably from the 1DRG curve, which results in their crossing. For a grater $c = 1.8$ value, the Cahn dependence has the greater early-stage value, $c\alpha_s\tau^4$, and a slower 1D asymptotics, than the curve in Fig. (12a); these facts result in its cross with the present-model curve for gamma DF.

Fig. (13b) shows that the extended JM distribution (which is a wide one) makes the $n(\tau)$ dependence for $\alpha_s = 0.1$ more smooth. Similarly to Fig. (9d), the distributions deflect the curves from the 1DRG one for identical spheres at a large $\alpha_s$ value. The curve for the above uniform DF is given for comparison; the similarity between the two wide distributions – extended JM and uniform – is obvious.

Fig. (14a) shows the $\ln(-\ln Q(\tau))$ -vs.- $\ln \tau$ plots corresponding to Fig. (13b) for a large $\alpha_s$. Differently from the Cahn curve consisting of the two parts corresponding to $n = 4$ and $n = 1$, the present-model curves have an additional (third) part corresponding to the 1DRG. The latter is a bend ending the curve. This bend is steep for identical spheres, whereas size distributions decrease its slope. However, it becomes steep again for wide distributions, with simultaneous slight lowering the curve as a whole. To study in more detail the dependence of the bend slope on the distribution width, the normal DF

$$f(R_0) = \frac{1}{\sigma\sqrt{2\pi}} \mathrm{e}^{-\frac{(R_0 - \bar{R}_0)^2}{2\sigma^2}} \qquad (57)$$

with different rms values $\sigma_r = \sigma / \bar{R}_0$ was utilized. The result is shown in Fig. (14b).

Just such form of the $\ln(-\ln Q(\tau))$ -vs.- $\ln \tau$ curve is observed in experimental studies of the crystallization of bulk metallic glasses [50]. This is a strong argument in favor of the grain structure of the latter and nucleation at grain boundaries. The polycluster model of amorphous solids was offered by Bakai [51] and employed for calculating the kinetics of crystallization of a bulk metallic glass within the Cahn model in ref. [32]. Intercluster boundaries are preferable places for nucleation. The kinetics of solidification of a supercooled liquid with the competitive formation and growth of crystalline and non-crystalline nuclei [52] was studied in ref. [53]; this process results in the formation of a polycluster amorphous solid.

## 7. Diffusional growth

The linear growth of nuclei was employed above for VF calculations, which corresponds to the interface-controlled mechanism of growth. Another important mechanism is the diffusion-limited growth occurring in multicomponent systems. Although the growth law does not change the general pattern of transformation considered above, it determines AE values. Therefore, the diffusional growth is briefly considered here as well.

The simplest form of diffusion-type law,

$$\frac{dR}{dt} = \frac{C}{2R}, \quad R(t', t) = \sqrt{C(t - t')} \tag{58}$$

is employed here for VF calculations; $C$ is a constant proportional to the diffusivity. The key quantities defined above have the following form now:

$$R_m(t) = \sqrt{Ct}, \quad r_0 = R_0 - \sqrt{Ct}, \quad t_m(r,t) = t - \frac{(R_0 - r)^2}{C}, \quad r_0' = \sqrt{Ct} - R_0,$$

$$t_m'(r,t) = t - \frac{(R_0 + r)^2}{C} \tag{59}$$

We introduce the dimensionless time $\tau = Ct / R_0^2$ and distance $x = r / R_0$ as well as the characteristic parameter $\alpha_s^{(d)} = \pi(I_s / C)R_0^4$. After the calculation of the integrals of Section 2, the following equations are obtained:

$$Q(\tau) = \begin{cases} \left(1 - \sqrt{\tau}\right)^3 + 3\int\limits_{1-\sqrt{\tau}}^{1} e^{-\alpha_s^{(d)}\varphi_1^{(d)}(x,\tau)} x^2 dx, & \tau < 1 \\ 3\left\{ \int\limits_0^{\sqrt{\tau}-1} e^{-\alpha_s^{(d)}\varphi_2^{(d)}(x,\tau)} x^2 dx + \int\limits_{\sqrt{\tau}-1}^{1} e^{-\alpha_s^{(d)}\varphi_1^{(d)}(x,\tau)} x^2 dx \right\}, & 1 \le \tau \le 4 \\ 3\int\limits_0^1 e^{-\alpha_s^{(d)}\varphi_2^{(d)}(x,\tau)} x^2 dx, & \tau > 4 \end{cases} \tag{60a}$$

where

$$\varphi_1^{(d)}(x,\tau) = \frac{\left[\tau - (1-x)^2\right]^2}{2x}, \quad \varphi_2^{(d)}(x,\tau) = 4(\tau - 1 - x^2) \tag{60b}$$

The corresponding KJMA equation (for bulk nucleation) is $X_K(\tau) = 1 - \exp[(8/5)\alpha_s^{(d)}\tau^{5/2}]$.

For instantaneous nucleation, the characteristic parameter is $\gamma_s = N_s S_0 = 4\pi N_s R_0^2$, as before. The VF is as follows:

$$Q(\tau) = \begin{cases} \left(1-\sqrt{\tau}\right)^3 + 3\int\limits_{1-\sqrt{\tau}}^{1} e^{-\gamma_s \varphi_0^{(d)}(x,\tau)}\, x^2 dx, & \tau < 1 \\[2mm] \left(\sqrt{\tau}-1\right)^3 e^{-\gamma_s} + 3\int\limits_{\sqrt{\tau}-1}^{1} e^{-\gamma_s \varphi_0^{(d)}(x,\tau)}\, x^2 dx, & 1 \le \tau \le 4 \\[2mm] e^{-\gamma_s}, & \tau > 4 \end{cases}$$

$$\varphi_0^{(d)}(x,\tau) = \frac{\tau - (1-x)^2}{4x} \tag{61}$$

For the surface transformation, one obtains

$$Q_s^{(cont)}(\tau) = \begin{cases} e^{-\alpha_s^{(d)}\tau^2/2}, & \tau \le 4 \\ e^{-4\alpha_s^{(d)}(\tau-2)}, & \tau > 4 \end{cases}, \quad Q_s^{(inst)}(\tau) = \begin{cases} e^{-\gamma_s \tau/4}, & \tau \le 4 \\ e^{-\gamma_s}, & \tau > 4 \end{cases} \tag{62}$$

The summand $(1-\sqrt{\tau})^3$ is the 1DRG VF for diffusional growth. Similarly to Eq. (24), the 1DRG AE is

$$n^{(1D)}(\tau) = -\frac{1}{2}\frac{\sqrt{\tau}}{(1-\sqrt{\tau})\ln(1-\sqrt{\tau})}, \quad n^{(1D)}(\tau) \to \begin{cases} 1/2, & \tau \to 0 \\ \infty, & \tau \to 1 \end{cases} \tag{63}$$

The diffusional-growth plots shown in Fig. (15) are in general similar to those considered above, but with half-integer AE values inherent in this (3D) case. The $n(\tau)$ dependence for a small $\alpha_s$ value varies between 2.5 and 1 for the reason described above.

Equations for the grain-boundary nucleated transformation can be derived as shown above. For more complex diffusional growth laws, a VF equation can be derived in the similar way. The actual growth equations have to take into account the depletion of any component around a nucleus, which stops its growth; this effect is especially important for the transformation of an isolated particle. Also, the simultaneous formation and growth of several phases with different compositions and the redistribution of components between them are possible. These processes as well as the possible crystallization-induced porosity (due to the difference in the densities of crystalline and amorphous phases) require the modification of the present VF equations as well.

## 8. Conclusions

1. VF equations, both nonisothermal and isothermal, have been obtained in the framework of Kolmogorov's method modified for the present problem. Characteristic parameters governing the transformation kinetics have been determined - $\alpha_s$ and $\gamma_s$ for continuous and instantaneous nucleation, respectively. The transformation process proceeds quite differently at small and large values of these parameters.

2. At small $\alpha_s(\gamma_s)$ values, the process is similar to that with bulk nucleation, accordingly, the AE changes from 4 to 1 (from 3 to 0) with time. A peculiarity of surface nucleation manifests itself just at large values of $\alpha_s(\gamma_s)$. In this case, the surface is covered by the crust of new phase at the early stage and further the one-dimensional radial growth of this crust occurs. As a consequence, the AE temporal behavior is non-monotonic; it has a minimum with subsequent sharp increase.

3. The kinetics of transformation of an ensemble of spherical particles is determined by the same parameters $\alpha_s$ and $\gamma_s$. The logarithmic normal and uniform size distributions employed for numerical examples stretch in time the VF and AE dependences for an ensemble of identical particles. At large $\gamma_s$ values, the logarithmic normal distribution with a large value of its width qualitatively changes the AE behavior which becomes monotonically decreasing with time.

4. A new VF equation derived for grain-boundary nucleated transformations shows that the kinetics of this process is governed by the same characteristic parameter $\alpha_s(\gamma_s)$ again. It qualitatively differs from the Cahn model kinetics for both small and large $\alpha_s$ values. At small $\alpha_s$s, the AE is near 4 for all times, as in the KJMA model, whereas the Cahn AE falls with time to 1. At large $\alpha_s$s, both the models give the same AE behavior at the initial stage of transformation, but quite different one at the late stage. Grain size distributions stretch VF and AE dependences in time, as in the ensemble of particles.

5. Being an origin of the AE temporal dependence, the logarithmic VF plots at large $\alpha_s$ values demonstrate a qualitative difference between the present and Cahn model as well; they drastically diverge at the late stage of the process. The plots of the present model end by a bend which is steep in the two opposite cases of very narrow and wide distributions. Experimental observations of such form of the logarithmic VF plots for the crystallization of bulk metallic glasses indicate to the polycluster structure of the latter and nucleation at intercluster boundaries.

**Appendix**

**1. Age distribution of grains in a nonisothermal case**

The mean volume of a $t'$-age grain at time $t$ is $V_{t'}(t) = dX_{t'}(t)/dN_{t'}$, where $dX_{t'}(t)$ is the VF of $t'$-age grains at time $t$ (the partial $t'$-age VF) and $dN_{t'} = Q(t')I(t')dt'$ is the number of $t'$-age grains. $I(t')$ is the bulk nucleation rate; the subscript "b" is omitted for brevity. The partial VF $dX_{t'}(t)$ can be obtained from the equation for the total VF $X(t)$ [2]

$$\frac{dX(t)}{dt} = Q(t)\frac{dY(t)}{dt}, \quad Y(t) = \int_0^t I(t')V(t',t)dt', \quad \frac{dY(t)}{dt} = \int_0^t I(t')\frac{\partial V(t',t)}{\partial t}dt' \tag{A1}$$

Substituting here $Q(t) = 1 - X(t)$ and integrating subject to the initial condition $X(0) = 0$, we arrive at Eq. (1a). For our purpose, however, we integrate it in the "general form":

$$X(t) = \int_0^t d\tau\, Q(\tau)\int_0^\tau I(t')\frac{\partial V(t',\tau)}{\partial \tau}dt' \tag{A2}$$

Changing the order of integration, we get

$$X(t) = \int_0^t dt' I(t') \int_{t'}^t Q(\tau)\frac{\partial V(t',\tau)}{\partial \tau}d\tau \equiv \int_0^t dt'\frac{dX_{t'}(t)}{dt'} \tag{A3}$$

from where

$$dX_{t'}(t) = I(t')dt' \int_{t'}^t Q(\tau)\frac{\partial V(t',\tau)}{\partial \tau}d\tau \tag{A4}$$

and

$$V_{t'}(t) = \frac{1}{Q(t')}\int_{t'}^t Q(\tau)\frac{\partial V(t',\tau)}{\partial \tau}d\tau \tag{A5}$$

$V_{t'}(t)$ is a monotonically decreasing function of $t'$, $dV_{t'}(t)/dt' < 0$, hence, there is the one-to-one correspondence between $t'$ and $V_{t'}(t)$. A basic equation for the DF is

$$f(V_{t'},t)dV_{t'} = -C(t)dN_{t'} = -C(t)Q(t')I(t')dt' \tag{A6}$$

where $C(t)$ is the normalizing coefficient; we put $C(t) = N^{-1}(t)$ for normalizing to unity,

$$N(t) = \int_0^t Q(t')I(t')dt' \tag{A7}$$

is the total number of grains at time $t$.

Thus,

$$f(V_{t'},t) = -\frac{1}{N(t)}\frac{Q(t')I(t')}{dV_{t'}(t)/dt'} \tag{A8}$$

$$\frac{dV_{t'}(t)}{dt'} = \frac{1}{Q(t')}\int_{t'}^{t}Q(\tau)\Phi(t',\tau)d\tau,$$

$$\Phi(t',\tau) \equiv \frac{\partial V(t',\tau)}{\partial \tau}\int_{0}^{t'}I(t'')\frac{\partial V(t'',t')}{\partial t'}dt'' + \frac{\partial^2 V(t',\tau)}{\partial t'\partial \tau} \tag{A9}$$

Finally,

$$f(V_{t'},t) = f(t'(V_{t'}),t) = -\frac{Q^2(t')I(t')}{N(t)}\left[\int_{t'}^{t}Q(\tau)\Phi(t',\tau)d\tau\right]^{-1} \tag{A10}$$

where the dependence $t'(V_{t'})$ is given by Eq. (A5). Eqs. (A10) and (A5) determine the DF $f(V_{t'},t)$ in a parametric form (the parameter is $t'$) and allow the study of its evolution in time as well as obtaining the final-state ($t = \infty$) DF in a nonisothermal case.

In the isothermal case, the following notations are used below: $k = (\pi/3)Iu^3$, $x^4 = kt'^4$, $y^4 = k\tau^4$, $z^4 = kt^4$, $v_1 = k^{1/4}I^{-1}$, $v_0 = 4\pi u^3 k^{-3/4}$; $v_0 = 12v_1$. The DF acquires the form

$$f(V_x,z) = f(x(V_x),z) = \left\{2v_0\left(\int_{0}^{z}\mathrm{e}^{-x^4}dx\right)\mathrm{e}^{2x^4}\int_{x}^{z}\mathrm{e}^{-y^4}\left[(y-x)-2x^3(y-x)^2\right]dy\right\}^{-1} \tag{A11}$$

$$V_x(z) = v_0\,\mathrm{e}^{x^4}\int_{x}^{z}\mathrm{e}^{-y^4}(y-x)^2dy \tag{A12}$$

These two equations determine the DF at time $z$ in a parametric form (the parameter is $x$).

The mean grain size in the final state, $z = \infty$, is

$$\overline{V}_0 = N^{-1}(\infty) = v_1/\zeta \qquad \zeta \equiv \int_{0}^{\infty}\mathrm{e}^{-x^4}dx = \frac{1}{4}\Gamma\left(\frac{1}{4}\right) \approx 0.906 \tag{A13}$$

Further, the dimensionless volume $v_x = V_x/\overline{V}_0$ and DF $f(v_x)$ are used:

$$v_x = 12\zeta\,\mathrm{e}^{x^4}\int_{x}^{\infty}\mathrm{e}^{-y^4}(y-x)^2dy \tag{A14}$$

$$f^{(JM)}(v_x) = f(x(v_x)) = \left\{24\zeta^2\,\mathrm{e}^{2x^4}\int_{x}^{\infty}\mathrm{e}^{-y^4}\left[(y-x)-2x^3(y-x)^2\right]dy\right\}^{-1} \tag{A15}$$

## 2. Extension of JM's approach

JM's approach replaces the actual distribution $f_x(\upsilon)$ of $x$-age grains by the $\delta$-distribution $f_x(\upsilon) = \delta(\upsilon - \upsilon_x)$. Therefore, the following step is to "broaden" this $\delta$-function, i.e. to allow some dispersion around the mean $\upsilon_x$. The normal distribution

$$f_x(\upsilon) = \frac{1}{\sigma_x \sqrt{2\pi}} \exp\left[ -\frac{(\upsilon - \upsilon_x)^2}{2\sigma_x^2} \right] \tag{A16}$$

with $\sigma_x = \xi_x \upsilon_x$ seems to be natural for this purpose; it goes to the $\delta$-distribution at $\sigma_x \to 0$. Here, $\xi_x$ is generally some function of $x$.

The increase in the volume of a $t'$-age grain at time $\tau$ is $dV^{(gr)}(t', \tau) = P(t', \tau)dV(t', \tau)$, $dV(t', \tau) = [\partial V(t', \tau)/\partial \tau]d\tau$, where $P(t', \tau)$ is the part of the spherical layer $dV(t', \tau)$ lying in the untransformed volume. Due to impingements with surrounding grains, $dV^{(gr)}(t', \tau)$ is a random quantity; the grain volume $V^{(gr)}(t', \infty)$ in the final state, as an integral, is the sum of great number of these small quantities. If the quantities $dV^{(gr)}(t', \tau)$ and $dV^{(gr)}(t', \tau + \Delta\tau)$ separated by a finite time interval $\Delta\tau$ can be thought nearly independent, then the central limit theorem under weak dependence can justify the normal distribution; however, this issue requires a more rigorous study.

The extended DF in the isothermal case has the form

$$f^{(ex)}(\upsilon) = \frac{\upsilon_1^{-1} \int_0^\infty f_x(\upsilon) Q(x) dx}{N(\infty)} = \zeta^{-1} \int_0^\infty f_x(\upsilon) e^{-x^4} dx \tag{A17}$$

The radius distribution is $f(r) = 3r^2 f(r^3)$, $r = R_0/\overline{R}_0$, $\overline{R}_0 = (3/4\pi)^{1/3} \overline{V}_0^{1/3}$.

It is worth noting that the variance $\sigma_x^2$ can be calculated by a probabilistic method without resort to a DF [41, 24]. The auxiliary DF $f_x^*(\upsilon)$ is used for this purpose, which is the probability for a random point to fall into the grain of volume $\upsilon$ in the $x$-population. According to the geometric definition of probability, this is

$$f_x^* = \frac{\upsilon N_x f_x(\upsilon) d\upsilon}{\upsilon_x^{(tot)}} = \frac{\upsilon}{\upsilon_x} f_x(\upsilon) \tag{A18}$$

where $N_x$ and $\upsilon_x^{(tot)}$ are the number of grains in the $x$-population and their total volume; $\upsilon_x = \upsilon_x^{(tot)}/N_x$. The mean for the function $f_x^*(\upsilon)$ is $\overline{\upsilon}^* = \overline{\upsilon^2}/\upsilon_x$, hence, $\sigma_x^2 = \overline{\upsilon^2} - \upsilon_x^2 = \overline{\upsilon}^* \upsilon_x - \upsilon_x^2$ and $\xi_x = (\overline{\upsilon}^*/\upsilon_x - 1)^{1/2}$. The procedure for calculating the function $\overline{\upsilon}^*(x)$ is described in ref. [24].

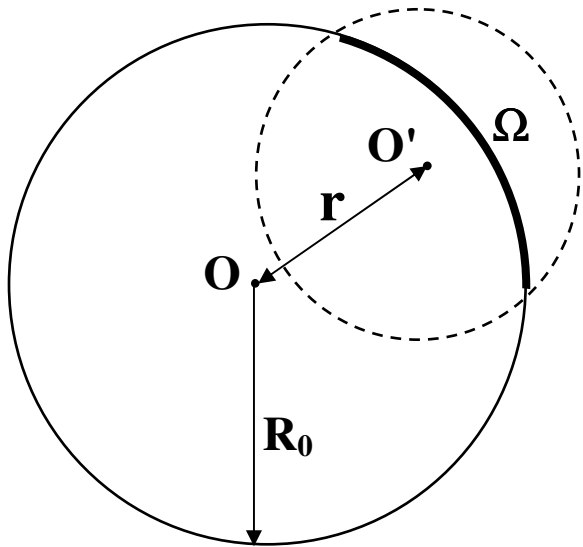

Fig. 1. Spherical particle (solid) and the CR for the point $O'$ (dashed). The CR encloses the part of particle surface (bold) with area $\Omega(r;t',t) = S(r;t',t)$.

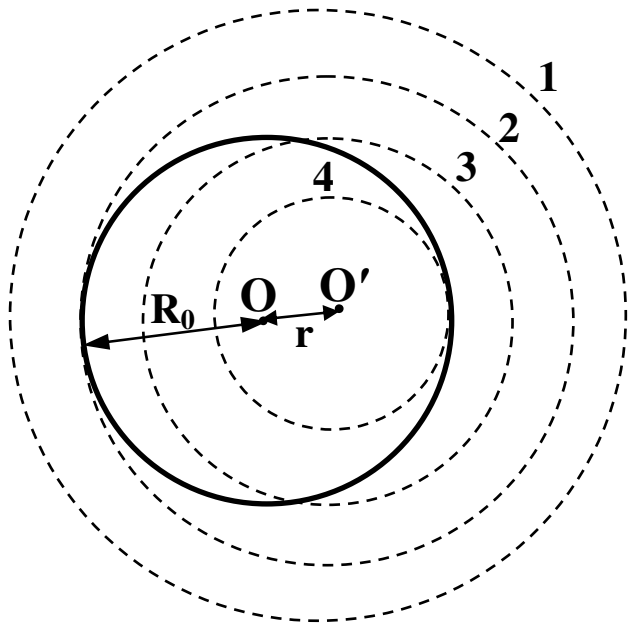

Fig. 2. Positions of the CR boundary at different times $t'$: (1) $0 \leq t' < t'_m(r,t)$; (2) $t' = t'_m(r,t)$; (3) $t'_m(r,t) < t' < t_m(r,t)$; (4) $t' = t_m(r,t)$.

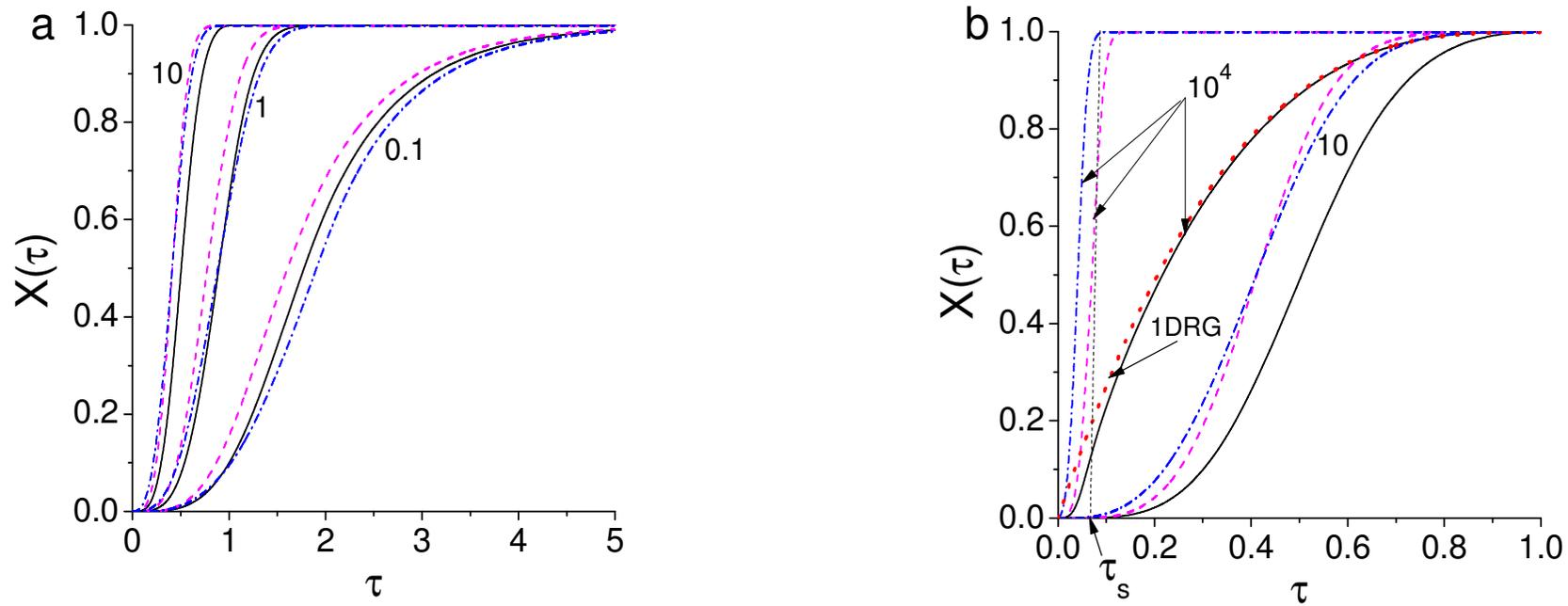

Fig. 3. Volume fractions for surface (solid) and bulk [11] (dashed) nucleation as well as the fraction of surface transformed (dash-dotted) at small (a) and large (b) values of the characteristic parameter $\alpha_s$ shown at the curves. The 1DRG curve $X^{(1D)}(\tau)$, Eq. (23), is shown by dotted line in Fig. (b); $\tau_s$ is the time of surface transformation.

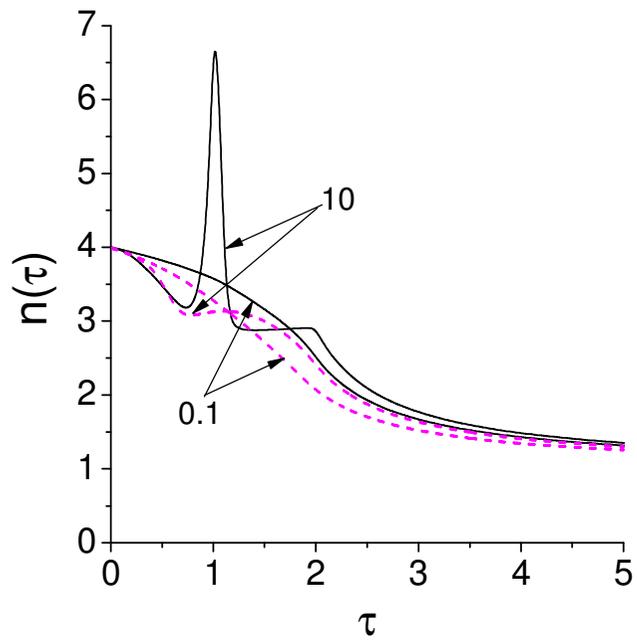

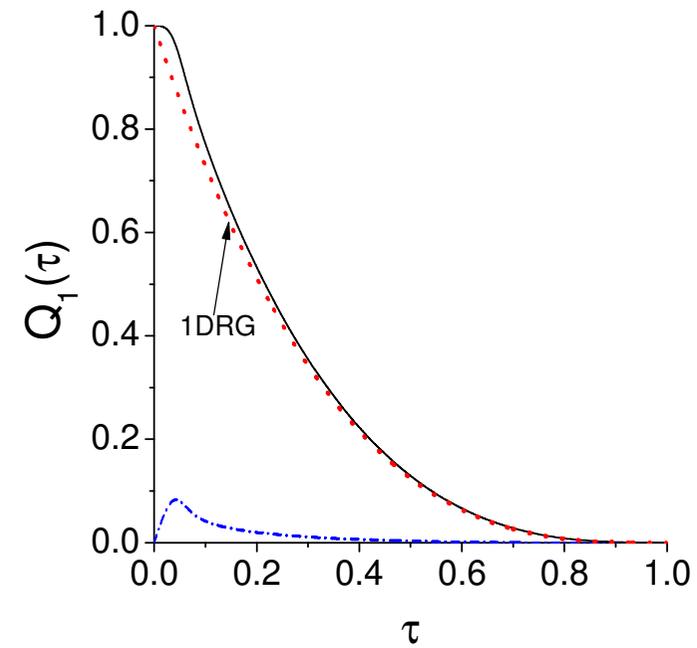

Fig. 4. AE as a function of time for surface (solid) and bulk (dashed) nucleation for the two values of $\alpha_s$ shown at the curves.

Fig. 5. VF $Q_1(\tau)$, Eq. (15a), (solid) and its constituents – the first summand (dotted) and the second one (dash-dotted) for $\alpha_s = 10^4$.

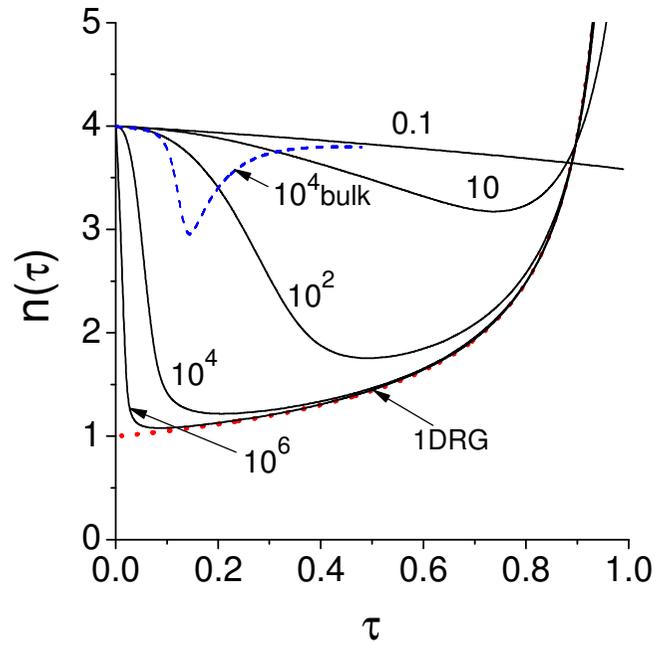

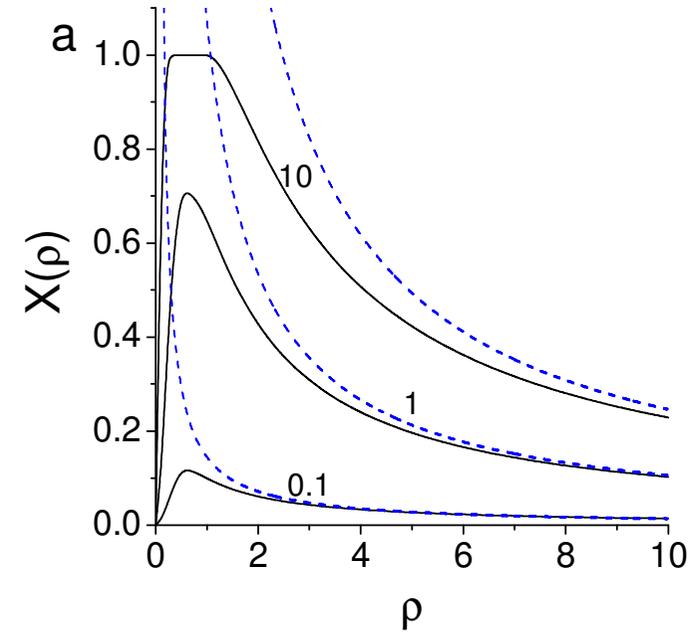

Fig. 6. AE for different values of $\alpha_s$ shown at the curves and the

1DRG curve $n^{(1D)}(\tau)$ (dotted). Dashed line is the AE for bulk nucleation

[11] at $\alpha_s = 10^4$ ( $\alpha_b = 3\alpha_s$ ).

Fig. 7a.

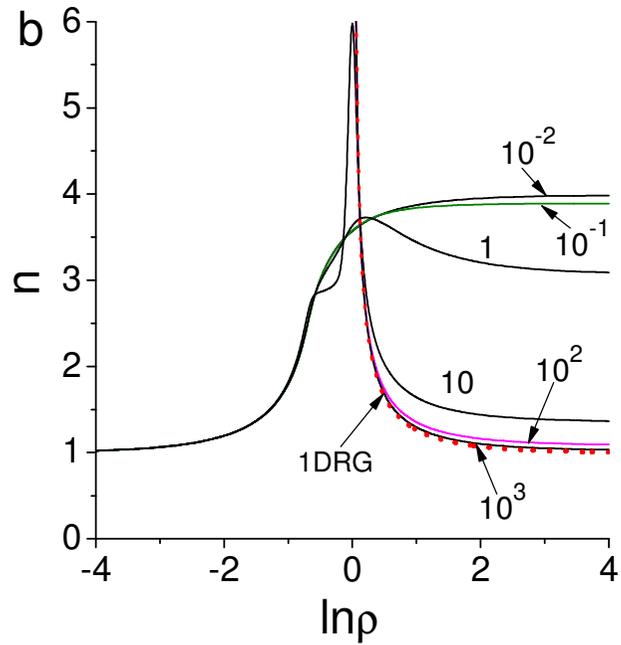

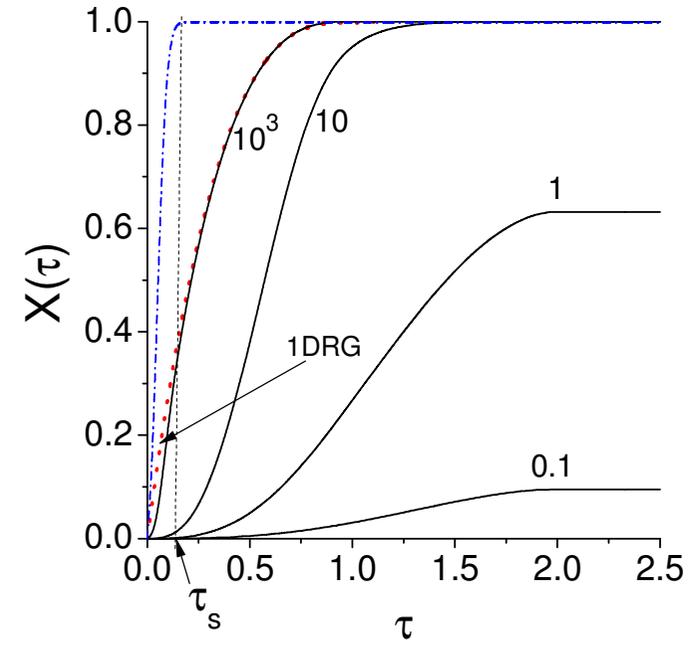

Fig. 7. (a) VF $X(\rho)$ (solid) with its approximation by Eq. (28b) (dashed) and (b) AE as a function of particle radius for different values of the characteristic parameter $\beta_s$ shown at the curves; the plots for $\beta_s = 10^2$ and $10^3$ are shown for $\rho > 1$.

Fig. 8. VF $X(\tau)$ for the case of instantaneous nucleation and different values of the characteristic parameter $\gamma_s$ shown at the curves. Dash-dotted curve is the surface transformation, Eq. 32, for $\gamma_s = 10^3$.

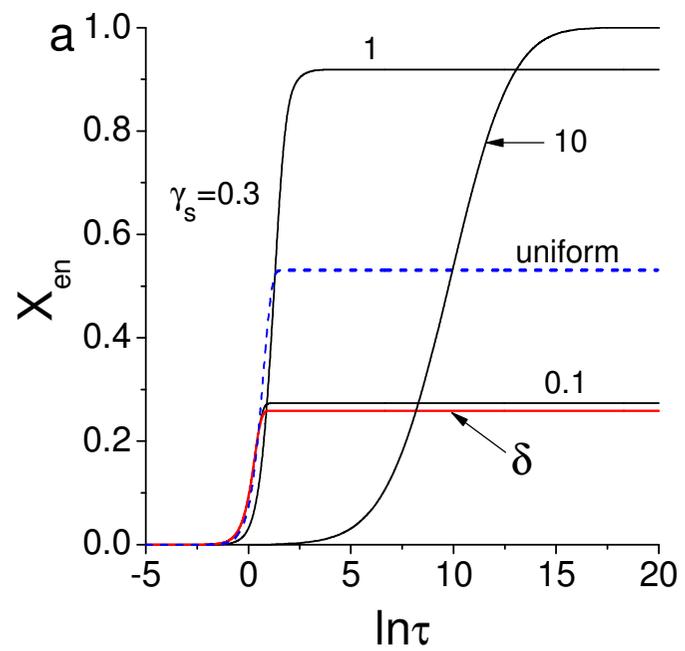

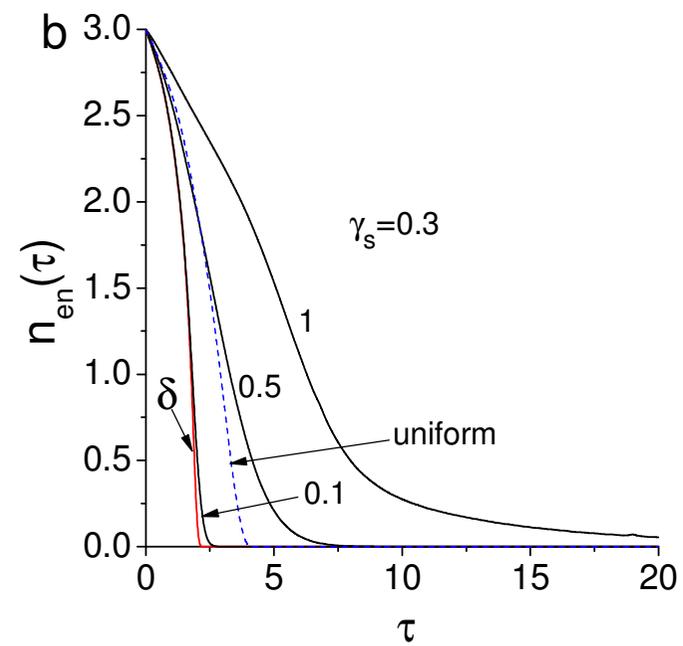

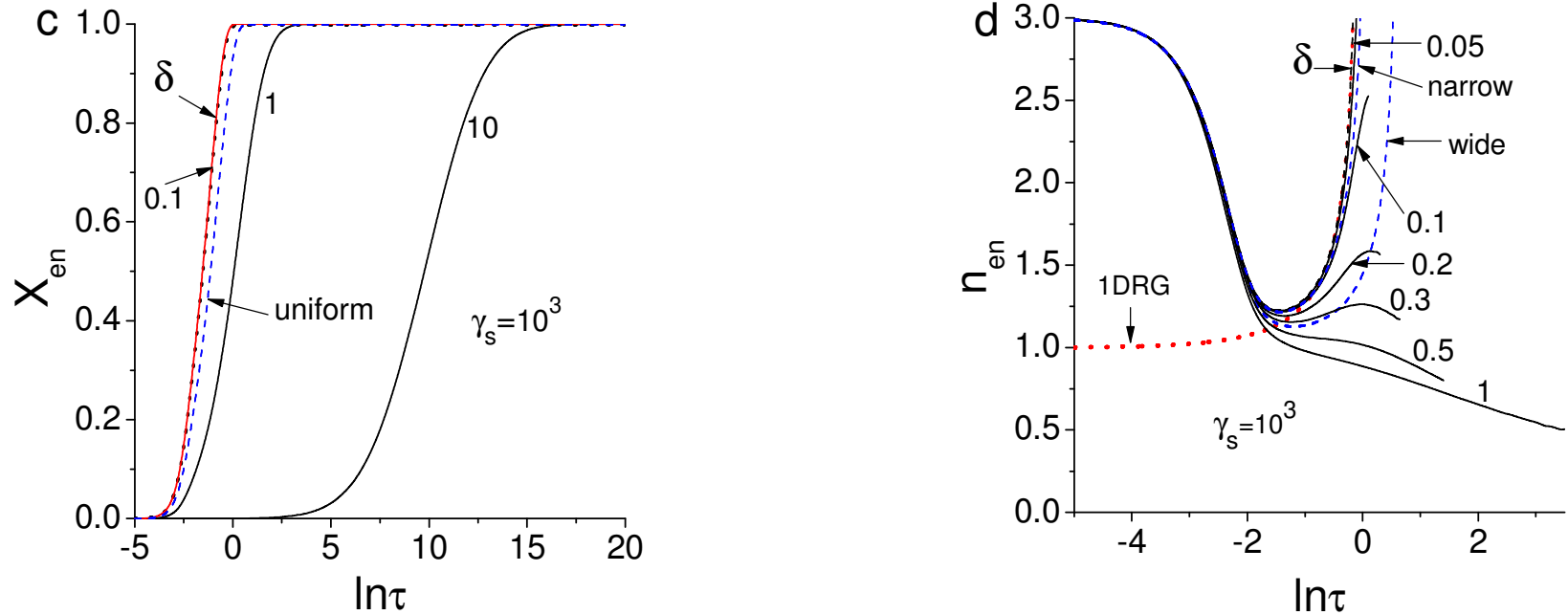

Fig. 9. VFs and AEs for the ensemble of particles with logarithmic normal and uniform size distributions for small (a, b) and large (c, d) values of the characteristic parameter $\gamma_s$ of instantaneous nucleation. The dependences are given for different values of the rms $\sigma_r$ shown at the curves. The plots for the ensemble of identical spheres of radius $\overline{R}_0$ are denoted by the symbol $\delta$ ($\delta$-shaped distribution). The 1GRG curve in Fig. (d) plots Eq. (24); dashed lines ibid are for the narrow and wide uniform distributions described in the text (in other figures, "uniform" relates to the wide one).

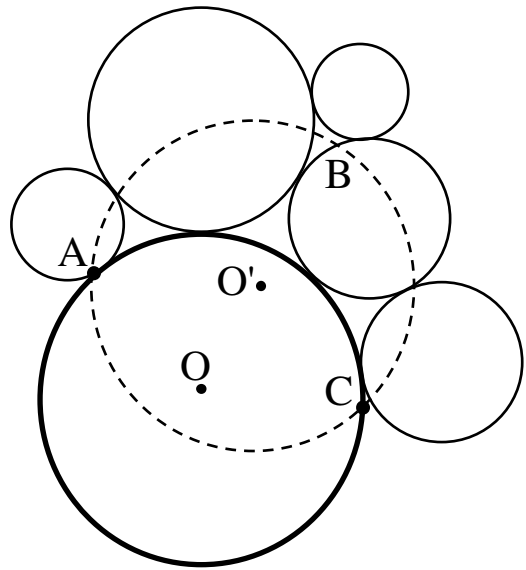

Fig. 10. Fragment of the network of grain boundaries in the present model. External boundaries for the given particle (bold) are within the segment ABC of the CR for the point $O'$ (dashed).

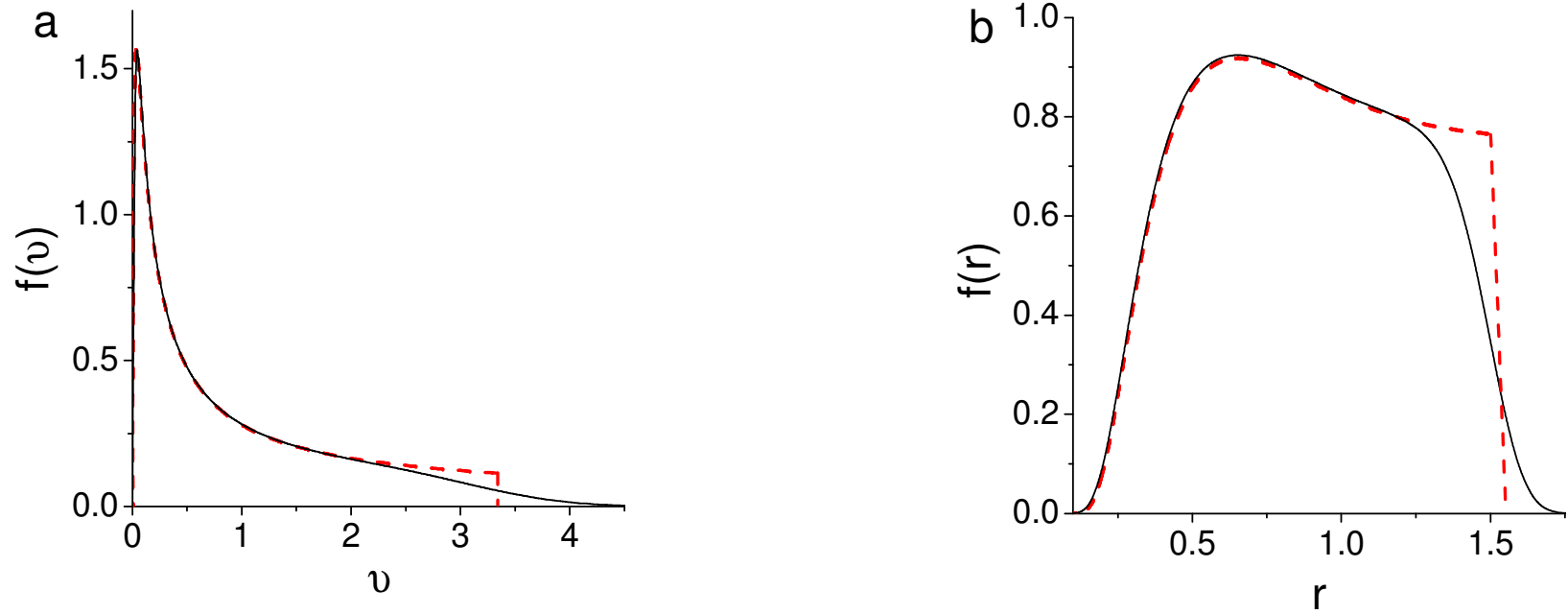

Fig. 11. (a) JM's (dashed), Eqs. (A14) and (A15), and extended JM's (solid), Eqs. (A17) and (A16) with $\xi_x = 0.2$, volume DFs. (b) Radius DFs corresponding to Fig. (a), JM's (dashed) and extended JM's (solid).

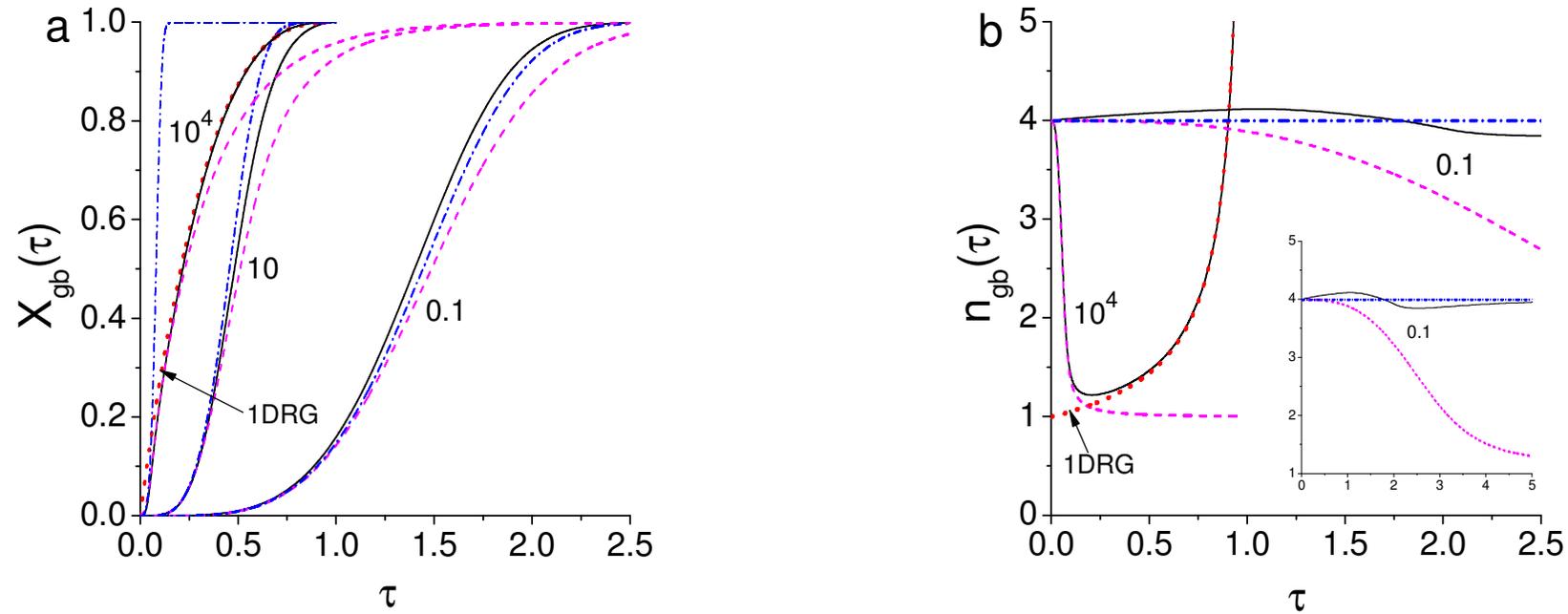

Fig. 12. VFs (a) and AEs (b) for grain-boundary continuous nucleation in the case of identical spherical grains with $c = 1.6$ at small and large values of the characteristic parameter $\alpha_s$ shown at the curves. Solid, dashed and dash-dotted lines relate to the present, Cahn, and KJMA models, respectively.

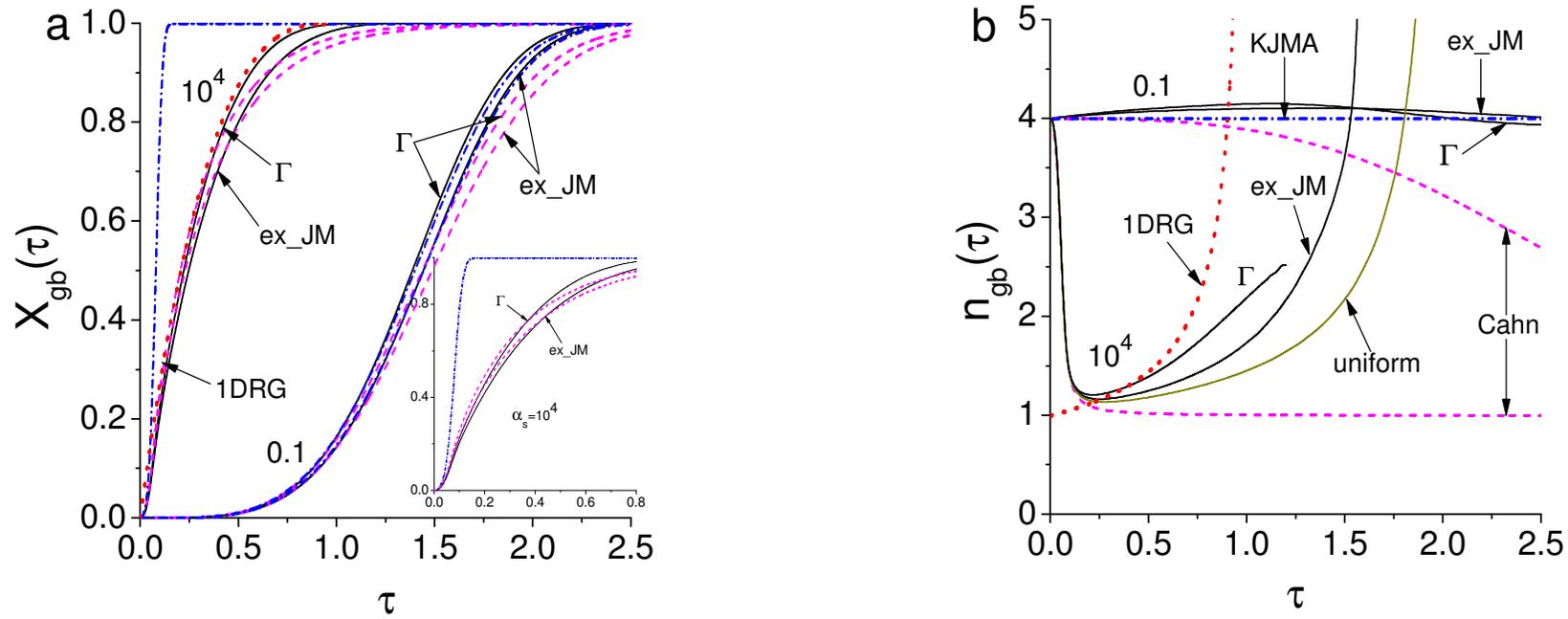

Fig. 13. VFs (a) and AEs (b) for grain-boundary continuous nucleation in the case of grain structure with the extended JM and gamma size distributions of grains at small and large values of the characteristic parameter $\alpha_s$ shown at the curves. Solid, dashed and dash-dotted lines relate to the present, Cahn, and KJMA models, respectively. The 1DRG curves are given for the system of identical spheres, Fig. (12).

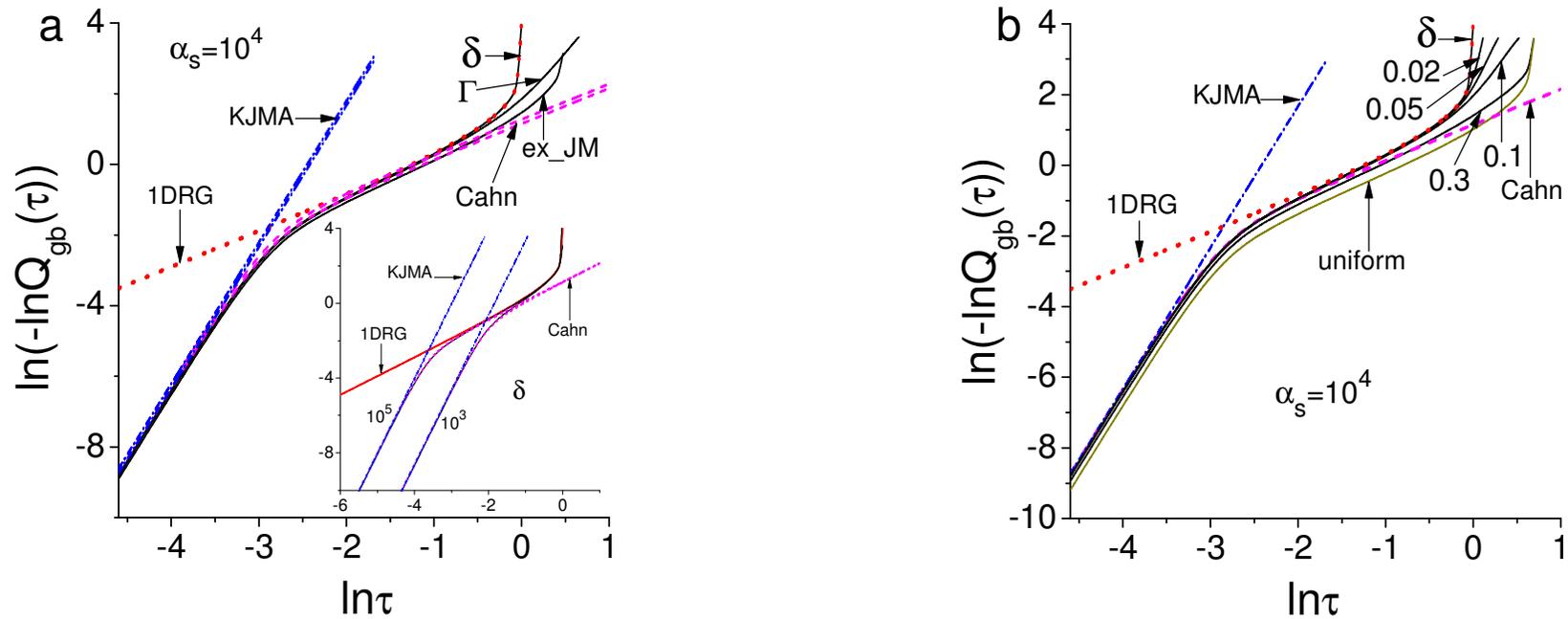

Fig. 14. Dependence $\ln(-\ln(Q(\tau)))$ vs. $\ln \tau$ for the grain-boundary nucleated transformation in comparison with the KJMA and Cahn models at the large value of $\alpha_s$. Fig (a) presents plots for the extended JM and gamma distributions; the inset shows plots for the grain structure composed of identical spheres for $\alpha_s = 10^3$ and $10^5$. Fig (b) shows plots for the normal DF, Eq. (57), with the rms values shown at the curves.

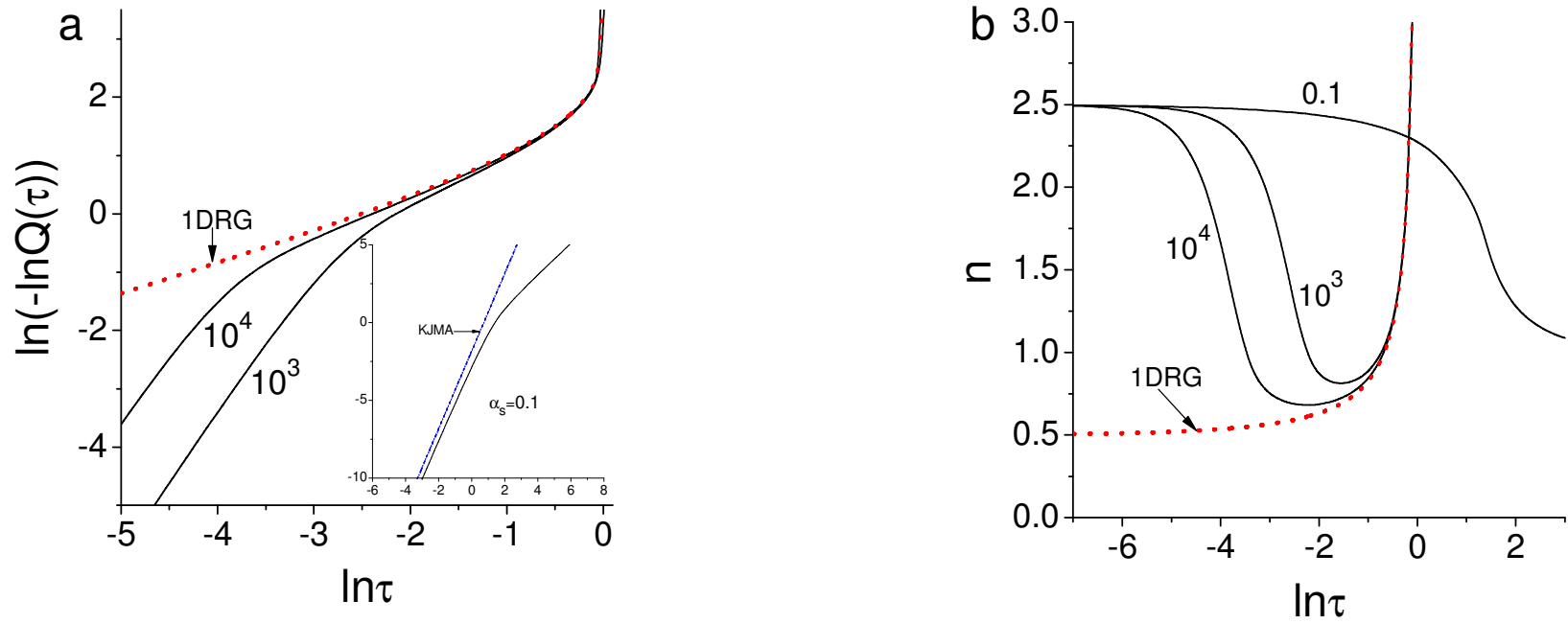

Fig. 15. Dependence $\ln(-\ln(Q(\tau)))$ vs. $\ln\tau$ and the corresponding AE for the transformation of a spherical particle in the case of diffusional growth for $\alpha_s$ values shown at the curves. The 1DRG curve in Fig. (b) plots Eq. (63).